\PassOptionsToPackage{pdfversion=2.0,pdfa,bookmarks,bookmarksnumbered=true,bookmarksopen=true,bookmarksopenlevel=0,ocgcolorlinks}{hyperref}
\documentclass[sigconf,nonacm]{acmart}

\usepackage{algorithm}
\usepackage[noend]{algpseudocode}
\usepackage{enumitem}
\usepackage{bm} %
\usepackage{booktabs} %
\usepackage{bold-extra} %
\usepackage[font={bf}, tableposition=top]{caption} %
\usepackage{graphicx} %
\usepackage{hyphenat} %
\usepackage{mathtools} %

\usepackage{microtype} %
\usepackage{multirow} %
\usepackage{subcaption} %
\usepackage{siunitx} %
\usepackage{xfrac} %
\usepackage{xspace} %
\usepackage{cleveref} %
\crefname{algorithm}{Alg.}{Algs.}
\Crefname{algorithm}{Algorithm}{Algorithms}
\crefname{appendix}{App.}{App.}
\Crefname{appendix}{Appendix}{Appendices}
\crefname{corollary}{Corol.}{Corolls.}
\Crefname{corollary}{Corollary}{Corollaries}
\crefname{conjecture}{Conjecture}{Conjectures}
\Crefname{conjecture}{Conjecture}{Conjectures}
\crefformat{conjecture}{Conjecture~#2#1#3}
\crefmultiformat{conjecture}{Conjectures~#2#1#3}{\ and~#2#1#3}{, #2#1#3}{\ and~#2#1#3}
\crefname{definition}{Def.}{Defs.}
\Crefname{definition}{Definition}{Definition}
\crefname{equation}{Eq.}{Eqs.}
\Crefname{equation}{Equation}{Equations}
\crefname{fact}{Fact}{Facts }
\Crefname{fact}{Fact}{Facts}
\crefname{figure}{Fig.}{Figs.}
\Crefname{figure}{Figure}{Figures}
\crefname{lemma}{Lemma}{Lemmas}
\Crefname{lemma}{Lemma}{Lemmas}
\crefname{proposition}{Prop.}{Props.}
\Crefname{proposition}{Proposition}{Propositions}
\Crefname{section}{Section}{Sections}
\crefname{section}{Sect.}{Sect.}
\crefname{subsection}{Sect.}{Sect.}
\Crefname{subsection}{Section}{Sections}
\crefname{subsubsection}{Sect.}{Sect.}
\Crefname{subsubsection}{Section}{Sections}
\crefname{table}{Table}{Tables}
\Crefname{table}{Table}{Tables}
\crefname{theorem}{Thm.}{Thms.}
\Crefname{theorem}{Theorem}{Theorems}
\PassOptionsToPackage{hyphens}{url} %
\usepackage[bb=dsserif]{mathalpha}

\usepackage{xcolor}
\hypersetup{
  pdfapart=1,
  pdfaconformance=B,
colorlinks=true,
linkcolor={red!50!black},
filecolor={green!50!black},
citecolor={green!50!black},
urlcolor={blue!80!black},
}

\graphicspath{{figs}}

\newcommand{\reals}{\ensuremath{\mathbb{R}}\xspace}
\newcommand{\expectation}{\ensuremath{\mathbb{E}}\xspace}

\newcommand{\graphfamily}{\ensuremath{\mathcal{G}}\xspace}
\newcommand{\nodes}{\ensuremath{V}\xspace}

\newcommand{\edges}{\ensuremath{E}\xspace}

\newcommand{\agraph}{\ensuremath{x}\xspace}
\newcommand{\anothergraph}{\ensuremath{x'}\xspace}
\newcommand{\graphvar}{\ensuremath{X}\xspace}
\newcommand{\obsgraph}{\ensuremath{x^*}\xspace}
\newcommand{\stats}{\ensuremath{g}\xspace}
\newcommand{\statsdim}{\ensuremath{p}\xspace}
\newcommand{\obsstats}{\ensuremath{\mathbf{g}^*}\xspace}
\newcommand{\distance}{\ensuremath{d}\xspace}
\newcommand{\scalevec}{\ensuremath{\mathbf{s}}\xspace}
\newcommand{\scalei}{\ensuremath{{s}}\xspace}
\newcommand{\scalematrix}{\ensuremath{\mathbf{S}}\xspace}

\newcommand{\diagonal}{\ensuremath{\mathrm{diag}}\xspace}
\newcommand{\tolerancebandshort}{\ensuremath{\mathcal{B}}\xspace}
\newcommand{\toleranceband}{\ensuremath{\tolerancebandshort({\obsgraph,\varepsilon})}\xspace}
\newcommand{\distribution}{\ensuremath{\mathcal{P}}\xspace}
\newcommand{\prob}{\ensuremath{\mathrm{Pr}}\xspace}
\newcommand{\probdistr}{\ensuremath{\prob_{_\distribution}}\xspace}
\newcommand{\onevec}{\ensuremath{\mathbb{1}}\xspace}
\newcommand{\intoleranceband}{\ensuremath{\onevec_{\tolerancebandshort}}\xspace}

\newcommand{\thetavec}{\ensuremath{\bm{\theta}}\xspace}
\newcommand{\entropy}{\ensuremath{\mathrm{H}}\xspace}
\newcommand{\dual}{\ensuremath{D}\xspace}
\newcommand{\iter}{\ensuremath{j}\xspace}
\newcommand{\proposal}{\ensuremath{q}\xspace}
\newcommand{\numsamples}{\ensuremath{M}\xspace}
\newcommand{\isample}{\ensuremath{i}\xspace}

\newcommand{\algopref}{BC\xspace}
\newcommand{\algopreflong}{Band-Con\-cen\-tra\-ted\xspace}
\newcommand{\algo}{\textsc{BaCon}\xspace}
\newcommand{\algolong}{\algopreflong ERGMs\xspace}

\setcopyright{acmlicensed}
\copyrightyear{2026}
\acmYear{2027}
\acmConference[WWW '27]{the ACM Web Conference 2027}{May 10-14, %
2027}{Dublin, Ireland}

\begin{document}

\title{\algo: \algopreflong Exponential Random Graph Models}

\author{Enrico Schiazzano}
\affiliation{%
  \institution{University of Turin}
  \city{Italy}
  \country{Italy}
}
\email{enrico.schiazzano@edu.unito.it}

\author{Giulia Preti}
 \orcid{0000-0002-2126-326X} %
\affiliation{%
  \institution{Intesa Sanpaolo Innovation Center}
  \city{Turin}
  \country{Italy}
}
\email{giulia.preti@intesasanpaolo.com}

\author{Aristides Gionis}
 \orcid{0000-0002-5211-112X} %
\affiliation{%
  \institution{KTH Royal Institute of Technology}
  \city{Stockholm}
  \country{Sweden}}
\email{argioni@kth.se}

\author{Matteo Riondato}
 \orcid{0000-0003-2523-4420} %
\affiliation{%
  \institution{Amherst College}
  \city{Amherst}
  \state{MA}
  \country{USA}}
\email{mriondato@amherst.edu}

\author{\texorpdfstring{\nohyphens{Gianmarco~De~Francisci~Morales}}{Gianmarco De Francisci Morales}}
 \orcid{0000-0002-2415-494X} %
\affiliation{%
 \institution{Intesa Sanpaolo Innovation Center}
 \city{Turin}
 \country{Italy}}
\email{gdfm@acm.org}

\begin{abstract}
Exponential-family random graph models (ERGMs) provide a parametric,
interpretable framework for modeling real graphs, including the Web and social
networks. Fitted ERGMs may suffer from \emph{degeneracy}: matching observed
graph statistics in expectation does not guarantee that the observed graph
and similar ones are highly probable under the fitted distribution. We introduce
\algolong (\algo), a framework that avoids degeneracy by imposing a
probabilistic constraint requiring at least $1-\delta$ of the distribution
probability mass to lie on the set of graphs with statistics within an
$\varepsilon$-tolerance band around the observed statistics. %
\algo's maximum-entropy distribution still belongs to the exponential family,
while including a factor checking for inclusion in the tolerance band. 
We derive the exact Metropolis–Hastings acceptance rule for this target, and use
a practical stochastic approximation procedure with a smoothed band membership during fitting.
Experiments on social networks across increasingly challenging model specifications show that \algo substantially increases the probability mass assigned to graphs
whose statistics are close to the observed ones, while remaining
estimable for specifications where standard ERGM estimation
fails. 
These results demonstrate that probability
mass constraints are a principled way to mitigate degeneracy while retaining
the benefits of exponential-family graph models.
\end{abstract}



\maketitle

\section{Introduction}\label{sec:intro}

Graphs are a fundamental representation of social, information, and interaction systems on the Web.
Statistical graph models aim to capture the structural features of an observed graph while generating plausible alternative realizations.
Exponential-family random graph models (ERGMs) provide a principled framework for this task by defining a probability distribution over graphs in terms of user-specified graph statistics~\citep{frank1986markov,wasserman1996logit}.
These statistics can encode properties such as degree heterogeneity, assortative mixing, and transitivity, while the corresponding parameters quantify how strongly these structures are favored by the model~\citep{snijders2006new,hunter2007curved}.
This combination of interpretable structural features and a generative
probabilistic model makes ERGMs a widely used tool for statistical graph analysis, with applications ranging from social networks~\citep{goodreau2009birds} to functional brain networks~\citep{simpson2011exponential}.
Because the normalizing constant of an ERGM involves a sum over the entire graph space, likelihood-based estimation and simulation typically rely on Markov chain Monte Carlo (MCMC) methods.

A persistent difficulty, however, is that an ERGM fitted to an observed graph
may not place substantial probability on graphs resembling that observation.
For some choices of statistics and parameters, the distribution becomes concentrated on extreme regions of graph space, often corresponding to nearly empty, nearly complete, or otherwise structurally implausible graphs~\citep{handcock2003assessing,snijders2006new}.
This phenomenon, known as \emph{degeneracy}, creates both statistical and computational problems.
Samples from the fitted model may have statistics far from those of the observed graph, while the Markov chain Monte Carlo (MCMC) algorithms used for estimation and simulation can struggle to move between separated high-probability regions~\citep{robins2007introduction}.
Degeneracy is closely connected to the geometry of exponential families and to sharp changes in the typical configurations favored by the model~\citep{fienberg2009geometry,chatterjee2013estimating}.
Consequently, a model may be mathematically well defined while providing a poor generative representation of the observed graph.  
For generative modeling of Web interaction graphs, degeneracy is
especially problematic because simulated graphs are intended to
provide plausible alternative realizations of an observed interaction
structure. Directed Web graphs also motivate dependence statistics for reciprocity, transitivity, and degree heterogeneity, making stable estimation of expressive graph models a practical modeling
challenge.

An extensive literature has therefore developed methods for constructing more stable ERGM specifications~\citep{snijders2006new,hunter2007curved,schweinberger2015local,fellows2017removing,blackburn2022practical}.
These methods have substantially improved the practical behavior of ERGMs.
However, they control the resulting distribution only indirectly, through
reweighted statistics, restrictions on dependence, penalties, or dispersion
bounds.
Consequently, the modeler cannot directly specify how much probability mass
should remain near the observed graph: whether the observation becomes
typical under the fitted model must instead be assessed after fitting, and may
depend sensitively on the chosen specification or tuning parameters.
The concentration of the distribution around the observed statistics is therefore an indirect consequence of these modeling choices rather than a property explicitly required by the model.

We take the
approach of specifying the desired behavior of the
distribution directly, enabling 
a natural middle ground between micro\-canonical and canonical graph ensembles~\citep{cimini2019statistical}.
Microcanonical models enforce selected statistics exactly, eliminating variation in those quantities, whereas canonical models constrain them only in expectation and place no direct bound on the probability of substantial deviations.
We instead allow the statistics to vary around their observed values while
requiring that large deviations occur with controlled probability.
Importantly, this modeling approach does not require abandoning the ERGM framework.
Under a maximum-entropy formulation, the resulting distribution retains an
exponential-family form with an additional band-indicator term and its associated calibration parameter, allowing inference and simulation to be carried out using modest modifications of standard MCMC procedures.

We formalize this idea through \emph{\algopreflong Exponential Random Graph
Models} (\algo ERGMs), a class of graph ensembles in which statistical fidelity
is imposed directly through a probability mass constraint.
For a graph \agraph, let $\stats(\agraph)\in\reals^\statsdim$ denote a vector of graph statistics, and $\obsstats \doteq \stats(\obsgraph)$ denote their values on the observed graph $\obsgraph$.
We use $\distance(\agraph,\obsgraph)$ to measure how far the statistics of $\agraph$ are from those of $\obsgraph$.
Given a tolerance $\varepsilon>0$ and a out-of-band probability $\delta\in(0,1)$, \algo requires an ensemble distribution \distribution such that
\begin{equation*}
    \probdistr\!\left[ \distance(\agraph,\obsgraph) \le \varepsilon \,\right] \;\ge\; 1-\delta\,,
\end{equation*}
together with the centering condition $\expectation[\stats(\agraph)] =
\obsstats$ (see \cref{sec:model} for details).
Here, $\varepsilon$ specifies the tolerated discrepancy in statistic space, while $\delta$ bounds how often that tolerance may be violated.
Among graph distributions satisfying these constraints, we consider a maximum-entropy formulation.
When uncertainty estimates for the observed statistics are available, the band may also be defined from those.
\algo therefore connects the concentration of a graph ensemble to quantities that can be specified before fitting the model.

Our contributions are threefold.
First, we formulate graph generation with an explicit $(\varepsilon,\delta)$ band constraint and derive its maximum-entropy solution, establishing its connection to an ERGM with an augmented sufficient statistic.
Second, we derive Metropolis--Hastings sampling for the target distribution and develop
a practical stochastic moment-matching procedure for fitting its
parameters, requiring only modest modifications of standard ERGM
machinery.
Third, we evaluate the approach on two empirical graphs and three model specifications exhibiting increasingly severe ERGM failure modes.
Experiments show that \algo improves the typicality of the observed graph while largely preserving the ordinary ERGM coefficients:
the fraction of simulated graphs whose statistics lie within the tolerance band increases by up to \(74\%\), while allowing model specifications where ERGM simply fails to converge.

\section{Related Work}\label{sec:relwork}

Exponential Random Graph Models (ERGMs) are distributions over the set of graphs
on a fixed set of identifiable vertices. They result from considering the
maximum entropy distribution under the constraint that the expectation of a
vector of statistics 
takes some desired
value, usually the one in an observed graph. They
have enjoyed immense success in modeling different types of graphs, and in studying
societal, biological, and physical phenomena. The
survey by \citet%
{cimini2019statistical} gives an
in-depth introduction to ERGMs. %

ERGMs are parametric models, where the probability assigned to a graph is
proportional to an exponential function of a linear combination of the vector
of statistics of the graph, weighted by the parameters of the model.
In practice, the parameters are fitted via Maximum Likelihood Estimation procedures.
Many ERGMs
exhibit \emph{degeneracy}~\citep{handcock2003assessing}: for most values of the
parameters, most of the probability mass is placed on a small region of the support, usually close to ``extreme'' graphs such as the clique and the empty
graph. The space of parameter values for which the probability mass is ``well
distributed'' on the support is often negligible.

Degeneracy arises when the fraction of graphs with the same vector of
statistics is, as a function of the vector, not log-concave~\citep{horvat2015reducing}. \Citet{schweinberger2011instability} highlights how
this behavior is inevitable when graph statistics have maxima that grow more
than quadratically with the number of vertices in a graph (e.g., the number of
triangles). The study of the theoretical consequences of degeneracy
has been extensive, and we refer the reader to the work by
\citet[Sect.\ 1 to 3, and references therein]{schweinberger2020exponential} for
a discussion of theoretical properties of ERGMs with respect to degeneracy.

This undesirable behavior is not a \emph{computational} issue; rather, it is a
model definition issue, but it affects the computational behavior of parameter
fitting algorithms such as Markov-Chain-Monte-Carlo Maximum Likelihood
Estimators (MCMC-MLE), and of algorithms to sample from the model.

Many solutions have been proposed to address degeneracy.
Mapping the original space to a log-concave one~\citep{horvat2015reducing} is
only possible in simple cases. Considering different families of statistics
depending on induced subgraph
counts~\citep{gunderson2020introducinggraphcumulantsvariance} is only reasonable
when these are indeed the statistics one wants to capture in the model. A more
general solution is the introduction of alternative models that are similar to
ERGMs, but address the root causes of degeneracy.

For example, ERGMs with additional structure~\citep[Sect.\ 4]{schweinberger2020exponential} 
impose a partial independence structure
on the model, allowing the probability of a graph to be expressed as a product
of multiple probabilities. For example, vertices may be partitioned into blocks and
factored over the set of pairs of blocks, i.e., every pair of blocks is
independent of the other pairs. The user must make a deliberate choice about
whether to impose such a structure, as it may or may not be appropriate for
their use case. %

A different class of models aims to eliminate degeneracy by acting on
the shape of the distribution over the set of graphs
via additional constraints.
In this sense,
our contribution belongs to this class of models. Curved
ERGMs~\citep{snijders2006new,hunter2007curved} introduce non-linear constraints
on the parameter space. A curved ERGM that uses the number of triangles in the
graph may 
have the value of an additional triangle decrease geometrically,
thus dampening the probabilities of graphs with a large number of triangles.
\Citet{fellows2017removing} propose ``tapered ERGMS,'' which introduce
constraints requiring the second moment of the difference between the expected
statistic and that of a graph to be bounded by a user-specified quantity.
This approach, while effective, introduces hyperparameters whose values are not easy to
choose a priori, although heuristics have been
proposed~\citep{blackburn2022practical}.
We introduce different constraints than any of these previously proposed models. Rather than
constraining (absolute) moments of the distribution, we require a user-specified
fraction of the probability mass to be assigned to the set of graphs whose
vector of statistics belongs to a user-specified domain.

A third approach %
is to restrict the support of the distribution.
\Citet{karwa2022dergms} propose to only consider graphs for which the maximum
$k$ such that the $k$-core of the graph is non-empty does not exceed a
user-specified value $k^*$. %
We take a more generic approach, where the tolerance band is specified with respect to the observed statistics and can act as a soft constraint.

\section{\algo ERGMs}\label{sec:model}

Let $\obsgraph=(\nodes,\edges)$ be the observed graph, and let \graphfamily\ be
the finite family of all graphs over the fixed set $\nodes$ of vertices.
For any graph $\agraph\in\graphfamily$, we consider the vector $\stats(\agraph)$
of $\statsdim$ sufficient statistics, $\stats:\graphfamily\to\reals^\statsdim$,
and we take $\obsstats \doteq \stats(\obsgraph)$.
To measure the similarity between a graph $\agraph\in\graphfamily$ and the
observed graph \obsgraph, we use the Euclidean distance on the scaled
statistics
\begin{equation} \label{eq:pac_distance}
  \distance(\agraph,\obsgraph) \doteq \|\, \scalematrix^{-1} (\stats(\agraph) - \obsstats) \,\|_{2}\,,
\end{equation}
where $\scalevec = (\scalei_1,\dots,\scalei_\statsdim)$ is a vector of scaling
factors, one per statistic, and $\scalematrix=\diagonal(\scalevec)$.
The user-specified vector $\scalevec$ can be used, for example, to give
different weights in the distance to the different statistics, or to normalize
the different statistics, preventing ones with larger scales (e.g., the
number of triangles) from dominating the distance over statistics with smaller
scales (e.g., a measure of transitivity). Distances other than the Euclidean can
be used. With $d$ as above, we can define an \emph{$\varepsilon$-tolerance
band} around $\obsgraph$.

\begin{definition}%
\label{def:pac_band}
Given a tolerance parameter~$\varepsilon>0$, the \emph{$\varepsilon$-tolerance
band} \toleranceband is the set of graphs in $\graphfamily$ whose distance from
\obsgraph is at most $\varepsilon$, that is,
\[
  \toleranceband \doteq
  \left\{\, \agraph  \in \graphfamily \mid \distance(\agraph,\obsgraph) \le
  \varepsilon \,\right\}.
\]
When \obsgraph and $\varepsilon$ are clear from context, we use \tolerancebandshort to denote~$\toleranceband$.
\end{definition}

We are interested in distributions over $\graphfamily$ where at least a
user-specified fraction of the probability mass is assigned to \toleranceband.

\begin{definition}[$(\varepsilon,\delta)$-\algopref distribution]%
\label{def:pac_distribution}
Let $\varepsilon \in \reals_{>0}$ and $\delta \in (0,1)$. 
A probability distribution \distribution on \graphfamily 
is an \emph{$(\varepsilon,\delta)$-\algopref distribution}, 
with respect to \obsgraph if it assigns probability mass at least $1-\delta$ to
the  $\varepsilon$-tolerance band \toleranceband, that is, 
\[
  \distribution(\toleranceband) \doteq 
  \probdistr \!\left[\, 
  \distance(\agraph,\obsgraph) \le \varepsilon \,\right] \;\ge\; 1-\delta \,.
\]
\end{definition}

We take the common approach of using the maximum-entropy
criterion~\citep{jaynes1957information,park2004statistical} to consider, for our
models, the least informative $(\varepsilon,\delta)$-\algopref distribution. The
maximization problem is the following, where $\mathcal{F}(\graphfamily)$ is the
family of functions from $\graphfamily$ to~$[0,1]$:
\begin{align}
  \max_{h \in \mathcal{F}(\graphfamily)}& \entropy(h) = -
  \sum_{\agraph\in\graphfamily} h(\agraph)\log h(\agraph)  \nonumber\\
  \text{such that}~& \sum_{\agraph\in\graphfamily} h(\agraph) \stats(\agraph) = \obsstats &
  \text{(centering)} \nonumber\\
  & \sum_{\agraph\in\graphfamily} h(\agraph) \intoleranceband(\agraph) \ge
  1-\delta &\text{(band concentration)}\label{eq:pac_primal}\\
 & \sum_{\agraph\in\graphfamily} h(\agraph) = 1 &\text{(probability
 axiom)}\nonumber
\end{align}

The space of feasible solutions is non-empty as it contains the
distribution that assigns all probability mass to $\obsgraph$. It is
convex and compact thanks to the properties of probability
distributions, the constraints are linear in $h$, and the entropy is strictly
concave. Therefore, Problem~\eqref{eq:pac_primal} is a concave maximization
problem with a unique optimal solution.
For the Lagrangian derivation below, we further assume that there exists a
strictly positive feasible distribution \(h^\circ\), i.e.,
\(h^\circ(\agraph)>0\) for every \(\agraph\in\graphfamily\), such that
\[
\sum_{\agraph} h^\circ(\agraph)\stats(\agraph)=\obsstats
\qquad\text{and}\qquad
\sum_{\agraph} h^\circ(\agraph)\intoleranceband(\agraph) \geq 1-\delta .
\]
Under this condition, strong duality holds, and the optimum admits
finite Lagrange multipliers~\citep[Ch.~5]{boyd2004convex}, yielding the
parametric form below.
If the condition does not hold, the maximum-entropy solution still exists on
the finite graph space, but it may lie on the boundary of the probability
simplex and require limiting, rather than finite, parameter values.
This issue %
is not specific to \algo,
but shared with standard ERGMs.

\begin{definition}[\algolong]\label{def:pac_ergm}
  Given $\tolerancebandshort$ and parameters $\thetavec\in\reals^\statsdim$ and
  $\nu\ge 0$, we call \emph{\algolong} the family of distributions of the form
  \begin{equation}
    \pi_{\thetavec,\nu}(\graphvar=\agraph) \doteq
    \frac{\exp\left\{\thetavec^{\top} \stats(\agraph) + \nu\,\intoleranceband(\agraph)\right\}}{Z(\thetavec,\nu)}\,,
    \label{eq:pac_model}
  \end{equation}
  where
  \begin{equation}
    Z(\thetavec,\nu) = \sum_{\agraph\in\graphfamily}
    \exp\left\{\thetavec^{\top} \stats(\agraph) + \nu\,\intoleranceband(\agraph)\right\}\,.
    \label{eq:pac_model_normalization}
  \end{equation}
\end{definition}
The target level $\delta$ does not enter the probability mass function
  directly: %
under the strict-feasibility condition above, the optimal parameters $(\thetavec^\star,\nu^\star)$ satisfy $\pi_{\thetavec^\star,\nu^\star}(\tolerancebandshort)\ge 1-\delta$.
\algolong are otherwise standard ERGMs with an additional sufficient statistic
for inclusion in \tolerancebandshort, boosting the probability of graphs in the
band by a factor $e^\nu$ over graphs outside the band.
The model recovers the standard ERGM for $\varepsilon \to
+\infty$ or $\delta \to 1$, and to the hard-constrained microcanonical ensemble
for $\varepsilon,\delta \to 0$.

\section{Fitting Algorithm}\label{sec:algorithm}
We now describe our approach to approximately finding the values for the parameters
$\thetavec$ and $\nu$ corresponding to the \algolong for the
observed graph $\obsgraph$ and tolerance band $\tolerancebandshort$.

The standard method of Lagrangian multipliers~\citep[Ch. 5]{boyd2004convex}
tells us that these values are those that minimize the Lagrangian dual
\begin{equation}\label{eq:pac_dual}
\dual(\thetavec,\nu) \doteq \ln Z(\thetavec,\nu) - \thetavec^\top \obsstats -
\nu(1-\delta) \enspace.
\end{equation}

The partial derivatives of $\dual(\thetavec,\nu)$ are
\begin{equation}\label{eq:pac_gradient:1}
\nabla_{\thetavec} \dual(\thetavec,\nu) = \expectation_{\thetavec,\nu}\!\left[\stats(\graphvar)\right] - \obsstats,
\end{equation}
and
\begin{equation}\label{eq:pac_gradient:2}
\frac{\partial}{\partial \nu}\dual(\thetavec,\nu) = \pi_{\thetavec,\nu}(\tolerancebandshort) - (1-\delta)\,.
\end{equation}
Each component of the gradient measures a discrepancy between the current
model and its target: the expected sufficient statistics vs.\ the
observed ones, and the achieved concentration vs.\ the required concentration.
As the partial derivatives depend on the partition function $Z$, whose
evaluation requires iterating over $\graphfamily$, computing  the partial
derivatives exactly is intractable. We develop a method to estimate them and the
minimizing parameters using Monte Carlo simulation and the Robbins-Monro
stochastic approximation~\citep{robbins1951stochastic}.

Our estimation procedure, summarized in \cref{alg:pac} in \cref{sec:algoapp}, combines
two nested steps: 
($i$)~we use Metropolis--Hastings~\citep{hastings1970monte} dynamics based
on the current parameters to obtain Monte Carlo estimates of the statistic and
concentration discrepancies; and 
($ii$)~we use Robbins-Monro stochastic approximation~\citep{robbins1951stochastic} to update~$\thetavec$ and~$\nu$ from these noisy estimates.

The procedure runs for $T$ Robbins-Monro iterations, chosen sufficiently large to allow the parameter trajectories to stabilize.
At each iteration $\iter$, we simulate $C$ independent Markov chains under the current parameters $(\thetavec^{(\iter)},\nu^{(\iter)})$.
Each chain is initialized at the observed graph $\obsgraph$, and an initial burn-in of $B$ Metropolis-Hastings steps is discarded. 
The burn-in is needed to move beyond the transient behavior induced by starting at $\obsgraph$: if it is too short, chains may remain artificially close to the observed graph, leading to an overestimate of the band concentration and consequently to an underestimate of $\nu$.
After burn-in, each chain contributes $\numsamples$ retained samples, separated by the sampling interval $L$. 

\paragraph{Sampling from the model.}
For fixed $(\thetavec,\nu)$, a graph \agraph is sampled from the distribution $\pi_{\thetavec,\nu}$ using Metropolis-Hastings. 
From \cref{eq:pac_model}, for two graphs \agraph and \anothergraph,
\begin{equation}
\frac{\pi_{\thetavec,\nu}(\anothergraph)}{\pi_{\thetavec,\nu}(\agraph)}
= \exp\left\{\thetavec^\top \Delta\, \stats + \nu\,\Delta\,\intoleranceband\right\}\,,
\label{eq:pac_rapporto}
\end{equation}
where
\begin{equation}
\Delta\,\stats = \stats(\anothergraph)-\stats(\agraph) ~\text{ and }~
\Delta\,\intoleranceband = \intoleranceband(\anothergraph)-\intoleranceband(\agraph)\,.
\end{equation}
Hence, given a proposal distribution $\proposal(\agraph\to\anothergraph)$, the Metropolis-Hastings acceptance probability is
\begin{equation}
\alpha(\agraph\to\anothergraph)
= \min\left\{1,\;
\exp\!\left(\thetavec^\top \Delta\, \stats + \nu\,\Delta\,\intoleranceband\right)
\frac{\proposal(\anothergraph\to\agraph)}{\proposal(\agraph\to\anothergraph)}
\right\}\,.
\label{eq:pac_mh}
\end{equation}

Notice that, since $\intoleranceband$ is binary,
\begin{equation}
\nu\,\Delta\intoleranceband =
\begin{cases}
\nu, & \text{if the move enters the band},\\
-\nu, & \text{if the move leaves the band},\\
0, & \text{otherwise}.
\end{cases}
\label{eq:pac_tre_casi}
\end{equation}
Thus, the additional term affects the Metropolis-Hastings ratio only when a proposal crosses the band boundary: entries are rewarded by a factor $e^\nu$, while exits are penalized by a factor $e^{-\nu}$.
For any proposal kernel that is irreducible on $\graphfamily$ and has
$\proposal(\agraph\!\to\!\anothergraph)>0$ only when
$\proposal(\anothergraph\!\to\!\agraph)>0$, the acceptance rule in
\cref{eq:pac_mh} satisfies detailed balance with respect to
$\pi_{\thetavec,\nu}$. Hence $\pi_{\thetavec,\nu}$ is stationary; with
aperiodicity, the chain converges to this target.

During estimation,
we replace the hard band indicator in the MCMC target by the smooth surrogate membership
\begin{equation}
\widetilde{\tolerancebandshort}_\rho(\obsgraph, \agraph, \varepsilon) = \sigma\!\left(\rho\,\frac{\varepsilon-\distance(\agraph,\obsgraph)}{\varepsilon}\right)\,,
\end{equation}
where $\sigma(z) = \frac{1}{1+\exp(-z)}$ is the logistic function and $\rho$ controls the sharpness
of the approximation~\citep{lenti2024likelihoodbased}.
The smooth membership provides a gradual increase in reward as a graph approaches the band boundary, improving mixing and stability.
The smooth surrogate is used only in the Metropolis-Hastings dynamics: the exact
indicator~$\intoleranceband$ is retained when estimating the concentration for the update of~$\nu$.
This surrogate is conservative, in the sense that the original model would give at least as much probability mass to graphs inside the band as the smoothed one.

\paragraph{Proposal distribution.}
We use a mixture proposal inspired by the neighborhood proposal of \texttt{ernm}~\citep{fellows2012exponential}.
At each Metropolis-Hastings step, we select with probability $\sfrac{1}{2}$ one between two moves.
The first is a tie/no-tie (TNT) move~\citep{morris2008specification}, which proposes either to remove an existing edge or to add an absent edge, each chosen uniformly at random. 
The second is a directed transitive move: a vertex $i$, one of its out-neighbors $j$, and one of $j$'s out-neighbors $k$ are sampled, and the edge $i \to k$ is toggled.

In our experiments, $\graphfamily$ is the set of directed simple graphs on the fixed vertex set, so the TNT move ensures irreducibility by allowing every edge to be added or removed.
The transitive move is added to improve exploration by directly proposing moves that affect triangles, which is particularly useful for specifications containing transitivity statistics.

\paragraph{Monte Carlo estimates.}
For chain $c$ at iteration $\iter$, let $\agraph^{(c,\iter)}_1$, $\ldots$, $\agraph^{(c,\iter)}_{\numsamples}$ denote the retained samples.
We compute the chain-wise estimates
\begin{equation}
\begin{aligned}
\widehat{\expectation}_{\thetavec,\nu}^{(c,\iter)}\!\left[\stats(\graphvar)\right]
&= \frac{1}{\numsamples}\sum_{\isample=1}^{\numsamples} \stats\!\left(\agraph_{\isample}^{(c,\iter)}\right) \,, \\
\widehat{\pi}_{\thetavec,\nu}^{(c,\iter)}(\tolerancebandshort)
&= \frac{1}{\numsamples}\sum_{\isample=1}^{\numsamples} \intoleranceband\!\left(\agraph_{\isample}^{(c,\iter)}\right)\,.
\end{aligned}
\label{eq:pac_mc_estimate}
\end{equation}

The sufficient statistics are aggregated component-wise using the median of the chain-wise means
, whereas concentration is aggregated using the mean
.
The median makes the statistic estimate less sensitive to individual chains that escape toward a distant mode.
The mean is used for concentration so that chains leaving the band contribute directly to the quantity controlled by $\nu$.

\paragraph{Updating the model parameters.}
The exact first-order conditions in \cref{eq:pac_gradient:1,eq:pac_gradient:2}
motivate the two discrepancy updates below. Because the samples are generated
with the smoothed dynamics and the sufficient statistics are aggregated robustly
across chains, these quantities are not exact stochastic gradients of
\cref{eq:pac_dual}. We use them to iteratively update the model parameters with a Robbins--Monro-style stochastic approximation~\citep{robbins1951stochastic} and decreasing step sizes $a_\iter = a_0 / (1+\iter c_\theta)$ and $b_\iter = b_0 / (1+\iter c_\nu)$.

\vspace{1mm}
\emph{Updating \texorpdfstring{$\thetavec$}{theta}.}
The update of \thetavec corrects the discrepancy between the observed and simulated sufficient statistics:
\begin{equation}
\thetavec^{(\iter+1)}
= \thetavec^{(\iter)} - a_\iter\,\scalematrix^{-2}
\left(\widehat{\expectation}_{\thetavec^{(\iter)},\nu^{(\iter)}}\!\left[\stats(\graphvar)\right]
- \obsstats\right),
\label{eq:pac_rm_theta}
\end{equation}
The matrix $\scalematrix$ is the same scale matrix used in \cref{eq:pac_distance}; thus, the term $\scalematrix^{-2}$ scales the Robbins-Monro update according to the variability of the sufficient statistics.
The terms in $\scalematrix$ are computed before the estimation of $\thetavec$ and $\nu$, by a band calibration procedure (\Cref{sec:results}).

To prevent unstable updates when the simulated distribution is far from the observed graph, we update $\thetavec$ only when the current estimated concentration is not lower than $0.5$.
Below this threshold, most sampled graphs lie far from $\obsgraph$, and the resulting statistic discrepancy may be dominated by extreme configurations. 
In such iterations, $\thetavec$ is kept fixed while $\nu$ continues to be updated.
The resulting increase in the band penalty drives the chains back toward the observed regime, after which the updates of $\thetavec$ resume.
Following the Polyak-Ruppert strategy~\citep{ruppert1988efficient, polyak1992acceleration}, the final estimate of \thetavec is obtained by averaging over the final $W$ iterations.

\vspace{1mm}
\emph{Updating \texorpdfstring{$\nu$}{nu}.}
The multiplier $\nu$ is updated from the difference between the achieved and the target concentration:
\begin{equation}
\nu^{(\iter+1)}
= \max\left\{0,\;\; \nu^{(\iter)} - b_\iter
\left(\widehat{\pi}_{\thetavec^{(\iter)},\nu^{(\iter)}}(\tolerancebandshort) - (1-\delta)\right)\right\}\,.
\label{eq:pac_rm_nu}
\end{equation}
Thus, $\nu$ increases when the estimated concentration is below the target and decreases when it is above the target.
Since the trajectory of $\nu$ can exhibit transient spikes, the final estimate of $\nu$ is obtained using the median over the same final window $W$ used for $\thetavec$.

The two recursions are performed within the same iteration using the same Monte Carlo samples, and provide a stochastic approximation to the minimizers of the dual from \cref{eq:pac_dual}.

\section{Empirical Evaluation}\label{sec:results}

Our evaluation addresses two questions: \textbf{(Q1)} Does \algo mitigate ERGM degeneracy? And \textbf{(Q2)} can the resulting stabilization make richer specifications usable, so that structural deficiencies can be modeled directly?
We consider three regimes of increasing difficulty for the standard ERGM: \textbf{(i)} estimation converges, but the fitted model is not centered on the observed statistics (\cref{sec:ukFaculty}); \textbf{(ii)} the model is correctly centered, but the observed graph may still be atypical because probability mass is excessively dispersed (\cref{sec:office_1}); and \textbf{(iii)} estimation fails altogether \cite{snijders2002markov} (\cref{sec:office_2}). These scenarios test, respectively, whether \algo can retain probability mass around the observed graph, avoid distorting an otherwise well-behaved ERGM while improving concentration, and make otherwise non-estimable specifications tractable.

\paragraph{Datasets and Model Specifications.}
To explore settings that are problematic for standard ERGMs, we use two directed social graphs previously studied in the ERGM literature (\cref{tbl:datasets}).
\texttt{ukFaculty}, first presented by \citet{nepusz2008fuzzy}, is a friendship graph among faculty members at a British university. An edge $(i,j)$ indicates that $i$ reports $j$ as a friend. Vertices belong to one of three schools (\texttt{GroupC}), allowing us to model group homophily and school-specific connectivity.
\texttt{office2005}~\cite{sailer2012social} records daily workplace interactions in a British university department. An edge $(i,j)$ indicates that $i$ reports a daily social interaction with $j$. The graph includes four dyadic covariates: perceived usefulness (\texttt{use}), indicators for membership in the same research team (\texttt{team}) and on the same building floor (\texttt{floor}), and walking distance between workstations (\texttt{metric}).

We consider three specifications, the last two drawn from prior analyses of \texttt{office2005}.

\textbf{\textsc{Spec}~1}, for \texttt{ukFaculty}, includes number of edges $|E|$, number of edges within the same school $H_{\mathrm{GroupC}}$ (school homophily), edges incident to each school $F^{\mathrm{GroupC}}_1$ and $F^{\mathrm{GroupC}}_2$, and geometrically weighted edgewise shared partners (GWESP) $T_\gamma$ (transitivity, with $\gamma=0.25$).
Note that GWESP is one of the most used alternative statistics proposed to capture transitivity while stabilizing ERGMs~\citep{snijders2006new}.

\textbf{\textsc{Spec}~2}, for \texttt{office2005}, is the model proposed by \citet{sailer2012social} and reproduced by \citet{clark2022comparing}.
Besides $|E|$ and transitivity $T_\gamma$ with $\gamma=0.5$, it models reciprocity (\texttt{rec}) and the four dyadic covariates $C_{\mathrm{use}}$, $C_{\mathrm{team}}$, $C_{\mathrm{floor}}$, $C_{\mathrm{metric}}$, each entering through its sum over observed edges.

\textbf{\textsc{Spec}~3} augments \textsc{Spec}~2 with geometrically weighted out- and in-degree terms, $D^{\mathrm{out}}_\phi$ and $D^{\mathrm{in}}_\phi$, with $\phi = 0.5$.
These terms summarize degree heterogeneity and follow the motivation of \citet{clark2022comparing}, who show that \textsc{Spec}~2 does not adequately reproduce the out-degree distribution and that adding degree terms leads the standard ERGM into a degenerate regime.

\begin{table}[!t]
    \centering
    \caption{Datasets and model specifications: vertices, edges, density, reciprocity, and sufficient statistics.}
    \label{tbl:datasets}
    \resizebox{\columnwidth}{!}{
        \begin{tabular}{lcccr@{\hspace{2em}}rl}
        \toprule
         \textbf{Dataset} & $|N|$ & $|E|$ & \textbf{Density} & \textbf{Recip.} & \multicolumn{2}{c}{\textbf{Specification}} \\
        \midrule
        \multirow{3}{*}{\texttt{ukFaculty}} & \multirow{3}{*}{79} & \multirow{3}{*}{781} & \multirow{3}{*}{0.127} & \multirow{3}{*}{229} & \multirow{3}{*}{1} & $|E|$, $H_{\mathrm{GroupC}}$ \\
        & & & & & & $F^{\mathrm{GroupC}}_1$ \\
        & & & & & & $F^{\mathrm{GroupC}}_2$, $T_\gamma$ \\
        \midrule
        \multirow{3}{*}{\texttt{office2005}} & \multirow{3}{*}{67} & \multirow{3}{*}{211} & \multirow{3}{*}{0.048} & \multirow{3}{*}{48} & \multirow{3}{*}{2} & $|E|$, $T_\gamma$, $\operatorname{rec}$ \\
        & & & & & & $C_{\mathrm{use}}$, $C_{\mathrm{team}}$ \\
        & & & & & & $C_{\mathrm{floor}}$, $C_{\mathrm{metric}}$ \\
        \midrule
        \multirow{2}{*}{\texttt{office2005}} & \multirow{2}{*}{67} & \multirow{2}{*}{211} & \multirow{2}{*}{0.048} & \multirow{2}{*}{48} & \multirow{2}{*}{3} &  \textsc{Spec}~2, \\
        & & & & & & $D^{\mathrm{out}}_\phi$, $D^{\mathrm{in}}_\phi$ \\
        \bottomrule
        \end{tabular}
    }
\end{table}

\paragraph{Experimental Setup.}
\Cref{tab:config-ext} in \cref{sec:configs} summarizes the main \algo parameters used throughout the experiments.
In all experiments we set $\delta=0.1$, i.e., a target concentration of $0.9$.
As the datasets do not come with reliability estimates for the measures, the band radius \(\varepsilon\) is calibrated, rather than fixed a priori.

We compare \algo with both standard ERGM and the Latent Order Logistic
(LOLOG) model~\citep{fellows2018new}.
LOLOG provides an alternative generative framework where potential edges are considered sequentially according to a latent order, and their formation probabilities are modeled through logistic functions of network change statistics.
This construction %
was proposed as a way to obtain more stable models
for specifications affected by ERGM degeneracy.

The standard ERGM is fitted by Monte Carlo maximum likelihood using the \texttt{ergm} package~\citep{hunter2008ergm}, with the same specifications and covariates as \algo, while LOLOG is fitted with the \texttt{lolog} package \citep{fellows2018new}.
On directed graphs, \texttt{lolog}'s $T_\gamma$ term counts cyclic shared partners (\(i\to j\to k\to i\)), whereas \texttt{ergm} and \algo use transitive shared partners (\(i\to k\to j\)). Thus, LOLOG parameter estimates are not directly comparable with those of ERGM and \algo, and its in-model evaluation uses the statistics of its own specification.

All experiments were run on an AMD Ryzen AI 7 350 CPU (16 threads) with 16\,GB of RAM, using Python~3.14 for \algo and R~4.5.3 (\texttt{ergm}~4.12.0, \texttt{lolog}~1.3.2) for the competing models.
The code is available at \url{https://anonymous.4open.science/r/pac_ergm-242F}.

\paragraph{Band Calibration.}
Defining the band \(\tolerancebandshort\) requires a scale vector \(\scalevec\) and a radius \(\varepsilon\), which we calibrate from \(2000\) graphs simulated from a dyad-independent reference model. 
The reference model %
preserves baseline effects due to density, node
attributes, and dyadic covariates, while excluding endogenous dependence
terms such as transitivity or degree effects.
This provides a simple notion of the natural variability of the
sufficient statistics without relying on the full ERGM.
The calibration procedure is not tied to the dyad-independent model: other reference distributions can be used.

We set each \(s_k\) to the empirical standard deviation of the corresponding statistic and \(\varepsilon\) as the \((1-\delta)\)-quantile of the standardized distances from the simulated mean. The resulting band therefore reflects the natural variability of the statistics under the reference model.
A sensitivity analysis in \cref{sec:res-eps} shows that varying \(\varepsilon\) mainly affects the concentration of the fitted distribution and the associated multiplier \(\nu\), while leaving coefficient estimates and out-of-model goodness of fit largely unchanged.

\paragraph{Evaluation Criteria.}
We evaluate the models along four increasingly stringent criteria.

\textbf{Estimability} records whether estimation converges, fails, or nominally converges while producing an unusable fit.
For \algo, the reported estimate $\hat{\thetavec}$ is the average $\thetavec$ over
the last $W$ iterations of a single run of \cref{alg:pac}. 
The values in
parentheses are ERGM-style conditional curvature diagnostics obtained from the
diagonal of $\mathrm{Cov}(\stats(X))^{-1}$, with the covariance estimated on
graphs simulated from the fitted model. They condition on the fitted multiplier
and calibrated band and do not account for their joint uncertainty; 
we therefore use them descriptively rather than for formal hypothesis tests.
To assess sensitivity to Monte Carlo randomness, we repeat each
\algo fit across five independent random seeds; \Cref{app:robustness} reports the resulting parameter
estimates and post-fit diagnostics.

\textbf{Centering} measures whether simulations from the fitted model match the observed sufficient statistics on average. Let $\widehat{\expectation}_{\thetavec,\nu}[\stats(X)]$ denote their empirical mean over simulated graphs. We quantify the
discrepancy using the standardized residual
\begin{equation}
  R =
  \left\lVert
  \frac{\widehat{\expectation}_{\thetavec,\nu}[\stats(X)]-\obsstats}{\scalevec}
  \right\rVert_2\,,
  \label{eq:residuo-std}
\end{equation}
where $\scalevec$ is the scale vector used by the \algopref distance. A small $R$ is necessary but not sufficient for a good fit: a distribution can be correctly centered while remaining highly dispersed or multimodal.

\textbf{Concentration} is measured by the resulting band concentration $\widehat{\pi}_{\thetavec,\nu}(\tolerancebandshort)$, computed as the fraction of simulated graphs inside the band.
The ideal \algo model requires $\pi_{\thetavec,\nu}(\tolerancebandshort)\geq1-\delta$.
For the fitted model, $\widehat{\pi}_{\thetavec,\nu}(\tolerancebandshort)$ is therefore a diagnostic rather than a guarantee of the approximate fitting procedure.

\textbf{Adequacy} assesses goodness of fit on quantities that are not included in the model specification. %
This provides an out-of-model assessment: unlike the coefficient comparison and the sufficient-statistic distributions, adequacy asks whether the fitted model also reproduces structural features that were not directly constrained. 

The Energy Score~\cite{gneiting2007strictly} provides a
common quantitative measure to assess fit by comparing models whose likelihoods are not directly comparable.
Given the standardized simulated statistics $\mathbf{h}_i=\stats(\agraph_i)/\scalevec$ and the standardized observation $\mathbf{h}^*=\obsstats/\scalevec$,
\begin{equation}
  \mathrm{ES}
  = \frac{1}{M}\sum_{i=1}^{M}\lVert \mathbf{h}_i-\mathbf{h}^*\rVert
  - \frac{1}{2M(M-1)}\sum_{i\neq j}\lVert \mathbf{h}_i-\mathbf{h}_j\rVert\,.
  \label{eq:energy-score}
\end{equation}
The first term measures deviation from the observation, while the second accounts for predictive dispersion; lower values indicate better agreement.
We use this score to assess both in-model and out-of-model statistics.

\subsection{Convergent but Miscentered ERGM}\label{sec:ukFaculty}
For \texttt{ukFaculty}, we estimate \algo in two stages: an initial phase with shorter chains to approach the solution, followed by a refinement phase with longer chains to accurately adapt \(\nu\) and stabilize band concentration.

\paragraph{Estimability.}
Both models are estimable. %
In \algo, \(\nu\) %
stabilizes at \(\hat{\nu}=2.50\). The mean band concentration over the final $30$ iterations is \(0.88\), close to the target \(1-\delta=0.9\).
This is an in-estimation diagnostic averaged over the final parameter iterates, distinct from the post-fit concentration reported below. %

\Cref{tab:spec-1-coeffs} reports the final coefficient estimates.
\algo and ERGM are closely aligned: the largest discrepancies, for $|E|$ and \(T_\gamma\), are approximately one tenth of the corresponding ERGM standard error.
Thus, the main difference between the two models lies in the additional band term \(\nu\,\intoleranceband\), and we can interpret the coefficients as in standard ERGMs.

\begin{table}[!t]
  \centering
  \caption{\textsc{Spec}~1. Estimated coefficients for ERGM and \algo. Standard errors in parentheses.}
  \label{tab:spec-1-coeffs}
  \resizebox{\columnwidth}{!}{
  \begin{tabular}{lr@{\hspace{5em}}r}
    \toprule
    & \textbf{ERGM} & \textbf{\algo} \\
    \midrule
    \(|E|\)                   & \(-6.030\) (0.220) & \(-6.006\) (0.195) \\
    \(H_{\mathrm{GroupC}}\)   & \(1.551\) (0.090)  & \(1.556\) (0.091)  \\
    \(F^{\mathrm{GroupC}}_1\) & \(0.000\) (0.048)  & \(0.002\) (0.046)  \\
    \(F^{\mathrm{GroupC}}_2\) & \(0.106\) (0.054)  & \(0.109\) (0.054)  \\
    \bottomrule
  \end{tabular}
  }
\end{table}

\begin{figure}[!t]
  \centering
  \includegraphics[width=\linewidth]{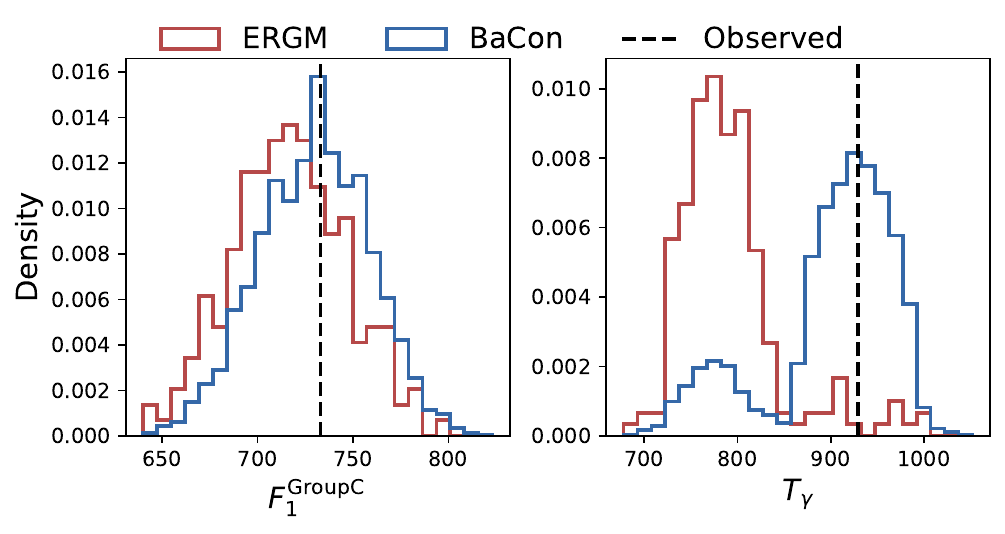}
    \caption{\textsc{Spec}~1. Distribution of $F^{\mathrm{GroupC}}_1$ and $T_\gamma$ under \algo and ERGM, and corresponding observed statistics.}
    \label{fig:uk-stats-compare}
\end{figure}

\begin{table}[!t]
  \centering
  \small
  \caption{\textsc{Spec}~1. In-model evaluation of \algo and ERGM. Best values are shown in bold.}
  \label{tab:spec-1-eval}
  \resizebox{.6\linewidth}{!}{
  \begin{tabular}{lccc}
    \toprule
    & \multicolumn{3}{c}{\emph{in-model evaluation}} \\
    \cmidrule(lr){2-4}
    & $R$ & \(\widehat{\pi}_{\thetavec,\nu}(\tolerancebandshort)\) & $\mathrm{ES}$ \\
    \midrule
    ERGM     &  7.26 & 0.07 & 5.83 \\
    \algo & \textbf{1.40} & \textbf{0.81} & \textbf{0.88} \\
    \bottomrule
  \end{tabular}
  }
\end{table}

\paragraph{Centering and Concentration.}
\Cref{fig:uk-stats-compare} compares the distributions of $F^{\mathrm{GroupC}}_1$ and $T_\gamma$ under the two fitted models; the remaining statistics are reported in \cref{sec:add-results}.
ERGM is not centered on the observed graph: $91\%$ of the simulated graphs fall in a sparse mode, with clearly fewer edges and lower $T_\gamma$ than observed, so that only $7\%$ fall inside the band.
By contrast, \algo shifts most of the probability mass toward the observed configuration: its upper mode is centered on $\obsstats$, and concentration rises to $81\%$.
Some graphs fall in the same sparse mode as under ERGM, keeping concentration below the nominal level $(1-\delta=0.9)$.
This behavior persists across independent runs: over five random seeds, the
post-fit concentration has mean $0.83$ (SD $0.10$), while the coefficient
estimates remain stable. The larger variability in concentration is
consistent with rare transitions to the remaining sparse mode
(\cref{app:robustness}).
The energy score decreases from $5.83$ for ERGM to $0.88$ for \algo (\cref{tab:spec-1-eval}).

\subsection{Centered but not Concentrated ERGM}\label{sec:office_1}
Following \citet{clark2022comparing}, we also compare against LOLOG.

\paragraph{Estimability.}
All three models converge.
\algo reaches 
mean concentration \(0.90\) in the final window, matching the target \(1-\delta\).

\algo and ERGM yield closely aligned coefficient estimates (\cref{tab:spec-2-coeffs}): the largest difference is about one third of an ERGM standard error, and all others are below one fifth.
The two models therefore lead to the same substantive conclusions, indicating that the \algopref constraint does not materially distort the ERGM parameter estimates while changing how probability mass is distributed around the observed graph.
Conversely, the coefficients estimated by LOLOG for most statistics are quite distant from the ERGM ones.

\begin{figure}[!t]
  \centering
  \includegraphics[width=\linewidth]{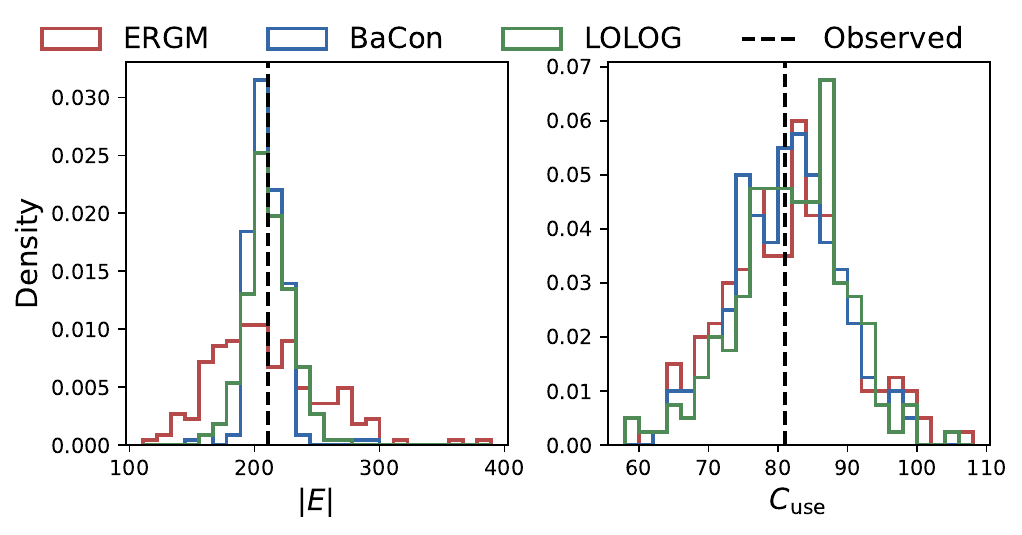}
  \caption{\textsc{Spec}~2. Distribution of $|E|$ and $C_\mathrm{use}$ under \algo, ERGM, and LOLOG, and corresponding observed statistics.}
  \label{fig:off-stats}
\end{figure}

\paragraph{Centering and Concentration.}
\Cref{fig:off-stats} compares $|E|$ and $C_\mathrm{use}$ under ERGM, \algo, and LOLOG; the remaining statistics are reported in \cref{sec:add-results}.
Unlike \texttt{ukFaculty}, the standard ERGM is well centered. %
However, its distribution is more dispersed, with only \(32\%\) of simulated graphs falling inside the band.

\algo remains comparably centered while concentrating substantially more probability mass around the observed graph: concentration increases to \(97\%\), and the energy score decreases from \(1.97\) to \(0.82\).
LOLOG provides an intermediate result, with concentration \(70\%\) and energy score \(0.97\).

\paragraph{Adequacy.}
Table~\ref{tab:spec-2-eval} compares the three models on the in-degree, out-degree, and shared-partner distributions using the energy score and \(\ell_1\) distance. 
ERGM and \algo perform similarly, as expected from their closely aligned coefficients.
Both reproduce the in-degree distribution reasonably well but fail on the out-degree, particularly the large number of observed vertices with zero out-degree, and underestimate the tail of the shared-partner distribution.
LOLOG performs similarly on in-degree but worse on out-degree and shared partners.
This out-degree mismatch motivates \textsc{Spec}~3.

\begin{table}[!t]
  \centering
  \caption{\textsc{Spec}~2. Estimated coefficients for ERGM, \algo, and LOLOG. Standard errors in parentheses.}
  \label{tab:spec-2-coeffs}
  \resizebox{\columnwidth}{!}{
  \begin{tabular}{lrr@{\hspace{2em}}r}
    \toprule
    & \textbf{ERGM} & \textbf{\algo} & \textbf{LOLOG} \\
    \midrule
    $|E|$ & $-3.358$ $(0.302)$ & $-3.362$ $(0.306)$ & $-2.176 (0.446)$\\
    $\operatorname{rec}$ & $0.564$ $(0.322)$ & $0.547$ $(0.324)$ & $1.887 (0.324)$\\
    $T_\gamma$ & $1.170$ $(0.102)$ & $1.188$ $(0.123)$ & $0.521 (0.119)$\\
    \texttt{$C_{\mathrm{use}}$} & $0.837$ $(0.140)$ & $0.828$ $(0.149)$ & $0.980 (0.147)$\\
    \texttt{$C_{\mathrm{team}}$} & $0.931$ $(0.125)$ & $0.892$ $(0.127)$ & $1.471 (0.185)$\\
    \texttt{$C_{\mathrm{floor}}$} & $0.038$ $(0.214)$ & $0.025$ $(0.202)$ & $0.049 (0.348)$\\
    \texttt{$C_{\mathrm{metric}}$} & $-0.043$ $(0.005)$ & $-0.042$ $(0.006)$ & $-0.063 (0.008)$\\
    \bottomrule
  \end{tabular}
  }
\end{table}

\begin{table}[!t]
  \centering
  \small
  \caption{\textsc{Spec}~2. In-model evaluation and out-of-model adequacy on the in-degree ($d^{\mathrm{in}}$), out-degree ($d^{\mathrm{out}}$), and shared-partner ($\mathrm{sp}$) distributions. Best values are shown in bold.}
  \label{tab:spec-2-eval}
  \resizebox{\linewidth}{!}{
  \begin{tabular}{lcccccccc}
    \toprule
    & \multicolumn{3}{c}{\emph{in-model evaluation}} & \multicolumn{5}{c}{\emph{out-of-model evaluation}} \\
    \cmidrule(lr){2-4} \cmidrule(lr){5-9}
    & \multirow{2}{*}{$R$} & \multirow{2}{*}{\(\widehat{\pi}_{\thetavec,\nu}(\tolerancebandshort)\)} & \multirow{2}{*}{$\mathrm{ES}$} & \multicolumn{3}{c}{$\mathrm{ES}$} & \multicolumn{2}{c}{$\ell_1$} \\
    \cmidrule(lr){5-7} \cmidrule(lr){8-9}
    & & & & $d^{\mathrm{in}}$ & $d^{\mathrm{out}}$ & $\mathrm{sp}$ & $d^{\mathrm{in}}$ & $d^{\mathrm{out}}$ \\
    \midrule
    ERGM     & 0.41 & 0.32 & 1.97 & $\mathbf{7.1}$ & $\mathbf{14.5}$ & $\mathbf{18.7}$ & $34.0$ & $48.1$ \\
    LOLOG    & \textbf{0.20} & 0.70 & 0.97 & $7.3$ & $25.4$ & $54.9$ & $33.1$ & $70.6$ \\
    \algo & 0.27 & \textbf{0.97} & \textbf{0.82} & $7.2$ & $14.6$ & $20.2$ & $\mathbf{31.8}$ & $\mathbf{47.0}$ \\
    \bottomrule
  \end{tabular}
  }
\end{table}

\subsection{Non-Estimable under Standard ERGM}\label{sec:office_2}

Following \citet{clark2022comparing}, \textsc{Spec}~3 augments \textsc{Spec}~2 with geometrically weighted in- and out-degree terms to address the degree-distribution mismatch. The corresponding standard ERGM is degenerate on this graph and cannot be estimated. For LOLOG, we follow \citet{clark2022comparing} and use second- and third-order in- and out-stars to address the same form of degree heterogeneity.

\paragraph{Estimability.}
The standard ERGM is not estimable, while both LOLOG and \algo converge.
Thus, \algo yields a valid estimate for a specification where standard ERGM estimation fails.

\Cref{tab:spec-3-coeffs} reports the \algo estimates.
The geometrically weighted out-degree coefficient \(D^{\mathrm{out}}_\phi\) is negative and large relative to its reported conditional curvature scale, whereas the in-degree coefficient \(D^{\mathrm{in}}_\phi\) is comparatively small.
This pattern is consistent with the motivation for the specification: degree heterogeneity is driven primarily by the outgoing degree distribution~\citep{clark2022comparing}.

\begin{figure}[!t]
  \centering
  \includegraphics[width=\linewidth]{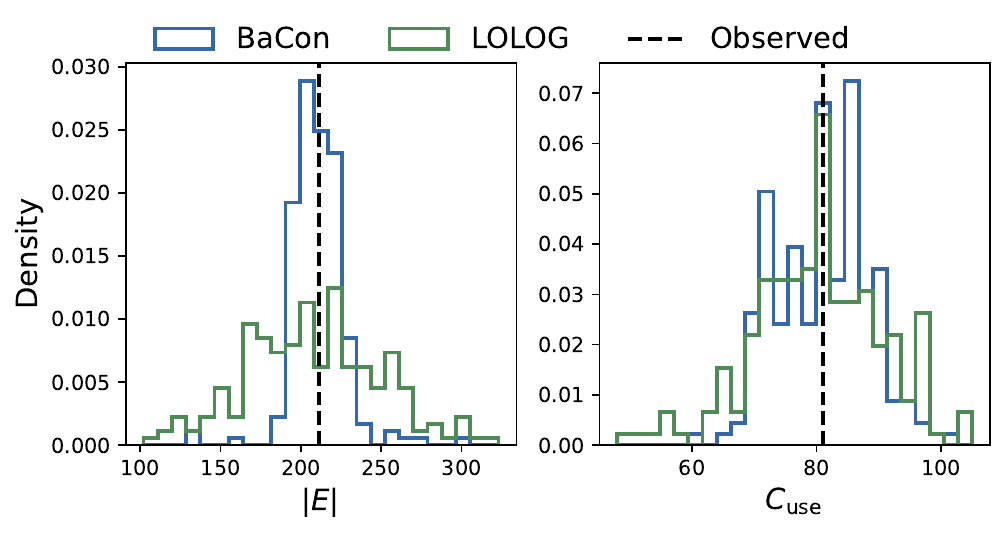}
  \caption{\textsc{Spec}~3. Distribution of $|E|$ and $C_\mathrm{use}$ under \algo and LOLOG, and corresponding observed statistics.}
  \label{fig:offgw-stats}
\end{figure}

\paragraph{Centering and Concentration.}
\Cref{fig:offgw-stats} shows the same statistics as \cref{fig:off-stats} after the addition of the two degree terms.
\algo keeps the same concentration around the observed values as in \textsc{Spec}~2 and achieves a band concentration of \(92\%\), close to the target \(1-\delta=0.9\). 

LOLOG with stars also reproduces these statistics on average, but its simulations become more dispersed (the standard deviation of $|E|$ grows from $18$ to $42$), yielding a band concentration of only \(19\%\).
As mentioned, LOLOG criteria are computed on the sufficient statistics of its own specification and should therefore be interpreted as indicative rather than directly comparable.

\paragraph{Adequacy.}
\Cref{tab:spec-3-eval} compares LOLOG and \algo on the in-degree, out-degree, and shared-partner distributions.
The main improvement is on out-degree: after introducing the degree terms, \algo reduces the energy score from \(14.6\) to \(5.5\) and the \(\ell_1\) distance from \(47.0\) to \(28.7\), substantially outperforming LOLOG with stars.
On in-degree, the two models perform similarly, while on shared partners \algo remains essentially unchanged from \textsc{Spec}~2 and slightly outperforms LOLOG.
This improvement should be attributed to the added degree statistics rather than to the band constraint alone. The contribution of \algo here is that this richer specification remains estimable, thus allowing the previously identified degree mismatch to be modeled directly.

\begin{table}[!t]
  \centering
  \small
  \caption{\textsc{Spec}~3. Estimated coefficients for \algo and LOLOG. Standard errors are in parentheses. For \algo, $z$-scores are also reported. (*) indicates statistics not included.}
  \label{tab:spec-3-coeffs}
  \begin{tabular*}{\columnwidth}{@{\extracolsep{\fill}}lrrrr@{}}
    \toprule
    & \multicolumn{2}{c}{\textbf{\algo}} & \textbf{LOLOG} \\
    \midrule
    & \textit{Estimate} & $z$ & \\
    \( |E| \) & \(-2.341\) \((0.406)\) & \(-5.77\) & -3.940 (0.815) \\
    \(\operatorname{rec}\) & \(0.812\) \((0.345)\) & \(2.35\) & 1.933 (0.768) \\
    \( T_\gamma \) & \(0.744\) \((0.152)\) & \(4.88\) & 0.291 (0.215) \\
    \( D^{\mathrm{out}}_\phi \) & \(-2.358\) \((0.444)\) & \(-5.31\) & (*) \\
    \( D^{\mathrm{in}}_\phi \) & \(0.316\) \((0.439)\) & \(0.72\) & (*) \\
    \( C_{\mathrm{use}} \) & \(0.809\) \((0.158)\) & \(5.13\) & 1.031 (0.200) \\
    \( C_{\mathrm{team}} \) & \(1.148\) \((0.180)\) & \(6.37\) & 2.249 (0.356) \\
    \( C_{\mathrm{floor}} \) & \(-0.057\) \((0.204)\) & \(-0.28\) & 0.197 (0.544)\\
    \( C_{\mathrm{metric}} \) & \(-0.0525\) \((0.007)\) & \(-8.06\) & -0.087 (0.014)\\
    \bottomrule
  \end{tabular*}
\end{table}

\begin{table}[!t]
  \centering
  \small
  \caption{\textsc{Spec}~3. In-model evaluation and out-of-model adequacy on the in-degree ($d^{\mathrm{in}}$), out-degree ($d^{\mathrm{out}}$), and shared-partner ($\mathrm{sp}$) distributions. Best values are shown in bold.}
  \label{tab:spec-3-eval}
  \resizebox{\linewidth}{!}{
  \begin{tabular}{lcccccccc}
    \toprule
    & \multicolumn{3}{c}{\emph{in-model evaluation}} & \multicolumn{5}{c}{\emph{out-of-model evaluation}} \\
    \cmidrule(lr){2-4} \cmidrule(lr){5-9}
    & \multirow{2}{*}{$R$} & \multirow{2}{*}{\(\widehat{\pi}_{\thetavec,\nu}(\tolerancebandshort)\)} & \multirow{2}{*}{$\mathrm{ES}$} & \multicolumn{3}{c}{$\mathrm{ES}$} & \multicolumn{2}{c}{$\ell_1$} \\
    \cmidrule(lr){5-7} \cmidrule(lr){8-9}
    & & & & $d^{\mathrm{in}}$ & $d^{\mathrm{out}}$ & $\mathrm{sp}$ & $d^{\mathrm{in}}$ & $d^{\mathrm{out}}$ \\
    \midrule
    LOLOG    & 1.20 & 0.19 & 3.32 & \textbf{5.2} & 10.7 & 23.3 & \textbf{29.7} & 43.6 \\
    \algo & \textbf{0.20} & \textbf{0.92} & \textbf{1.03} & 6.0 & \textbf{5.5} & \textbf{20.4} & 30.3 & \textbf{28.7} \\
    \bottomrule
  \end{tabular}
  }
\end{table}

\subsection{Scalability}\label{sec:scalability}
We assess the scalability of \algo on induced subgraphs of the Slashdot graph,\footnote{\url{https://snap.stanford.edu/data/soc-Slashdot0902.html}} with approximately $10^2$, $10^3$, $10^4$, and $10^5$ edges.
Subgraphs are obtained by breadth-first exploration from a randomly selected vertex.
On each subgraph, we fit a model with $|E|$,
\texttt{rec}, and $T_\gamma$ ($\gamma=0.5$), and simulate 10 graphs from the fitted model.

All runs use $T=300$, $C=8$, $\delta=0.1$, $B=10m$, $M=100$,
$L=m/10$, and $W=100$, where $m$ is the number of edges.
The band is calibrated from $200$ Bernoulli graphs, and estimation starts from the maximum pseudo-likelihood estimate (MPLE).
For simulation, we use a burn-in of $10m$ steps and retain one graph every $m$ steps.

\begin{figure}[!t]
\centering
\includegraphics[width=\linewidth]{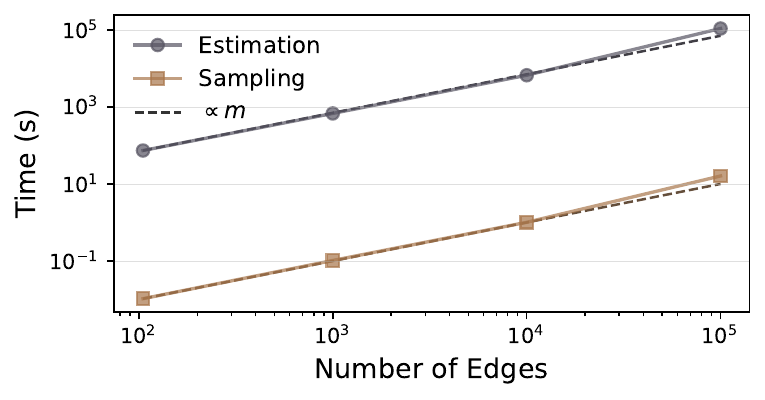}
\caption{Scalability of \algo on induced subgraphs of the Slashdot graph. Fitting time and simulation time per graph are reported against the number of edges \(m\); dashed lines indicate linear scaling \(\propto m\).}
\label{fig:scalability}
\end{figure}

\Cref{fig:scalability} reports the results on a log--log scale.
Both estimation and sampling times grow approximately linearly with the number of edges \(m\).
The increase from the \(10^4\)-edge to the \(10^5\)-edge subgraph is slightly superlinear
due to the larger degrees: after an edge toggle, updating $T_\gamma$ requires computing the shared partners of the affected vertices, whose cost increases with their degrees.

\section{Discussion}\label{sec:discussion}

Our results show that \algo mitigates different manifestations of ERGM degeneracy by explicitly controlling the probability mass assigned to graphs whose statistics are close to those observed.
On \texttt{ukFaculty}, where standard ERGM estimation produces overly sparse networks, \algo increases band concentration from $7\%$ to $81\%$ and reduces the energy score from $5.83$ to $0.88$.
On \texttt{office2005}, where the ERGM is centered but overly dispersed, concentration increases from $32\%$ to $97\%$.
Moreover, \algo remains estimable when adding geometrically weighted in- and out-degree statistics that cause standard MCMLE to fail, while largely preserving the original ERGM coefficients.
Although concentration alone does not guarantee better fit on properties outside the model specification, \algo enables richer specifications that can address these deficiencies, as demonstrated by the improved out-degree fit.

From a theoretical perspective, \algo provides a principled extension of maximum-entropy graph modeling starting from an explicit requirement on the probability mass assigned to a band of graph space.
This formulation provides a middle ground between canonical ensembles, which constrain statistics in expectation, and microcanonical ensembles, which enforce them exactly.
The resulting distribution retains the exponential-family form, with an additional sufficient statistic indicating membership in the tolerance band.
Importantly, the formulation can be connected to measurement error theory, allowing the tolerance $\varepsilon$ and failure probability $\delta$ to be informed by uncertainty estimates of the observed statistics, and thus providing a principled interpretation of these parameters.

Several limitations remain.
First, the concentration constraint applies only to the selected sufficient statistics, and its interpretation depends on their scaling and the tolerance region.
Second, in the absence of external uncertainty estimates, as in our experiments, the tolerance region depends on the chosen scaling and reference-model calibration.
Finally, despite the convexity of the dual problem, practical estimation relies on MCMC sampling and is therefore sensitive to mixing and convergence.
Future work should characterize the smoothing approximation, assess robustness across calibration choices, and evaluate generative quality on larger Web-derived graphs with complex structural dependencies.

\appendix

\section{Algorithmic Details}\label[appendix]{sec:algoapp}

\Cref{alg:pac} summarizes the estimation procedure from \cref{sec:algorithm}.

\begin{algorithm}[!h]
\small
\caption{\algo}
\label{alg:pac}
\begin{algorithmic}[1]
\Require Observed graph $\obsgraph$, Observed Statistics $\obsstats$
\Require Tolerances $\varepsilon, \delta$, Initial Parameters $\thetavec^{(0)}$ and $\nu^{(0)} \ge 0$
\Require Number of Chains $C$, Samples per Chain $\numsamples$
\Require Number of Iterations $T$, Burn-in $B$, Scale Matrix $\scalematrix$
\Require Polyak--Ruppert Window $W$, Sampling Interval $L$
\Ensure Estimates $\widehat{\thetavec}, \widehat{\nu}$
\For{$\iter=0,\dots,T-1$}
    \State $a_\iter\gets a_0/(1+\iter c_\theta)$ \Comment{Robbins-Monro step size for $\thetavec$}
    \State $b_\iter\gets b_0/(1+\iter c_\nu)$ \Comment{Robbins-Monro step size for $\nu$}
    \For{$c=1,\dots,C$} \Comment{Independent chains}
        \State Initialize $\agraph \gets \obsgraph$
        \For{$b=1,\dots,B$} \Comment{Burn-in interval}
            \State Propose $\agraph'\sim \proposal(\agraph\to\cdot)$
            \State Set $\agraph \gets \agraph'$ with probability $\alpha(\agraph\to\anothergraph)$ \Comment{Eq.\eqref{eq:pac_mh}}  \label{eq:smooth_accept_1}
        \EndFor
        \State $\graphfamily^{(c,\iter)} \gets$ empty list
        \For{$i=1,\dots,\numsamples$}
            \For{$\ell=1,\dots,L$}
                \State Propose $\agraph'\sim \proposal(\agraph\to\cdot)$
                \State Set $\agraph \gets \agraph'$ with probability $\alpha(\agraph\to\anothergraph)$ \label{eq:smooth_accept_2}
            \EndFor
            \State Add $\agraph$ to $\graphfamily^{(c,\iter)}$ \Comment{Sample every $L$ steps}
        \EndFor
        \State Compute $\widehat{\expectation}_{\thetavec,\nu}^{(c,\iter)}\!\left[\stats\right]$ and $\widehat{\pi}_{\thetavec,\nu}^{(c,\iter)}(\tolerancebandshort)$ from $\graphfamily^{(c,\iter)}$ \Comment{Eq.\eqref{eq:pac_mc_estimate}}
    \EndFor
    \State $\widehat{\expectation}_{\thetavec,\nu}^{(\iter)} \gets \operatorname{median}_{c = 1, \dots, C}\widehat{\expectation}_{\thetavec,\nu}^{(c,\iter)}$ \label{line:media_iter}
    \State $\widehat{\pi}_{\thetavec,\nu}^{(\iter)}(\tolerancebandshort) \gets \operatorname{mean}_{c = 1, \dots, C}\widehat{\pi}_{\thetavec,\nu}^{(c,\iter)}(\tolerancebandshort)$ \label{line:mean_iter}
    \If{$\widehat{\pi}_{\thetavec,\nu}^{(\iter)}(\tolerancebandshort) \geq 0.5$}
        \State $\thetavec^{(\iter+1)} \gets \thetavec^{(\iter)} + a_\iter\scalematrix^{-2} \left(\obsstats - \widehat{\expectation}_{\thetavec,\nu}^{(\iter)}\right)$ \Comment{Eq.\eqref{eq:pac_rm_theta}}
    \Else $\quad\thetavec^{(\iter+1)} \gets \thetavec^{(\iter)}$
    \EndIf
    \State $\nu^{(\iter+1)} \gets \max\left\{0, \nu^{(\iter)} + b_\iter\left((1-\delta)-\widehat{\pi}_{\thetavec,\nu}^{(\iter)}(\tolerancebandshort)\right)\right\}$ \Comment{Eq.\eqref{eq:pac_rm_nu}}
\EndFor
\State $\widehat{\thetavec} \gets \operatorname{mean}_{\iter = 1, \dots, W}\thetavec^{(T-\iter)}$; $\widehat{\nu} \gets \operatorname{median}_{\iter = 1, \dots, W} \nu^{(T-\iter)}$
\State \Return $\widehat{\thetavec},\widehat{\nu}$
\end{algorithmic}
\end{algorithm}

\section{\algo Parameter Configuration}\label[appendix]{sec:configs}

\Cref{tab:config-ext} reports the complete configurations used for
estimation and simulation in the three specifications. Most settings are
shared across experiments, e.g., the number of chains, Robbins-Monro
step sizes, logistic slope, and simulation burn-in. The main differences
concern the calibrated band radius, chain lengths, and sampling settings.
For \textsc{Spec}~1, entries of the form $x/y$ refer to the two estimation
stages, whereas \textsc{Spec}~3 starts from a preliminary estimate instead
of the MPLE.%

\begin{table}[!ht]
  \centering
  \small
  \caption{\algo settings across specifications.}%
  \label{tab:config-ext}
  \resizebox{\columnwidth}{!}{
  \begin{tabular}{lccc}
    \toprule
    & \textsc{Spec}~1 & \textsc{Spec}~2 & \textsc{Spec}~3 \\
    \midrule
    \multicolumn{4}{@{}l}{\emph{Estimation}} \\
    Tolerance $\varepsilon$                 & $3.47$      & $3.89$      & $4.31$ \\
    Initial $\thetavec$                     & MPLE\,/\,stage 1 & MPLE & prelim.\ est. \\
    Iterations $T$                        & $1000\,/\,100$ & $1000$ & $1000$ \\
    Number of chains $C$                  & $8$         & $8$         & $8$ \\
    Burn-in $B$                           & $200{,}000\,/\,2M$ & $500{,}000$ & $500{,}000$ \\
    Sampling interval $L$                        & $1000$      & $1000$      & $1000$ \\
    Samples per chain $M$                        & $250$       & $200$       & $200$ \\
    Coverage aggregation                         & median\,/\,mean & mean & mean \\
    Initial $\thetavec$ step $a_0$               & $0.3$       & $0.3$       & $0.3$ \\
    Decay $c_\theta$                 & $0.0005$    & $0.001$     & $0.001$ \\
    Initial $\nu$                                & $10\,/\,1$  & $10$        & $10$ \\
    Initial $\nu$ step $b_0$                     & $2.5$       & $2.5$       & $2.5$ \\
    Decay $c_\nu$                          & $0.005$     & $0.002$     & $0.002$ \\
    Logistic slope $\rho$                        & $16$        & $16$        & $16$ \\
    Polyak-Ruppert window $W$                   & $200\,/\,30$ & $200$      & $200$ \\
    \addlinespace
    \multicolumn{4}{@{}l}{\emph{Simulation}} \\
    Burn-in $B$                                       & $1M$        & $1M$        & $1M$ \\
    Number of samples                                 & $5000$      & $200$       & $200$ \\
    Sampling interval $L$                             & $10{,}000$  & $100{,}000$ & $100{,}000$ \\
    \bottomrule
  \end{tabular}
  }
\end{table}

\section{Sensitivity to the Band Radius}\label[appendix]{sec:res-eps}

We assess sensitivity to the band radius by re-estimating \textsc{Spec}~3 with
\(\varepsilon=k\hat\varepsilon\), for \(k\in\{0.5,0.75,1,1.5,3,4\}\), keeping the sufficient statistics, scale vector \(\scalevec\), and target level \(\delta\) unchanged. Here \(k=1\) corresponds to the calibrated radius \(\hat\varepsilon=4.31\).

\begin{table}[bt]
  \centering
  \small
  \caption{\textsc{Spec}~3. Sensitivity to \(\varepsilon=k\hat\varepsilon\). 
\(\widehat{\pi}_{k}(\tolerancebandshort)\) and
\(\widehat{\pi}_{1}(\tolerancebandshort)\) denote coverage of the fitted and
calibrated bands, respectively. ES is reported both in-model and out-of-model
for the in-degree, out-degree, and shared-partner distributions.}%
  \label{tab:eps-sens}
  \resizebox{\linewidth}{!}{
  \begin{tabular}{lcccc@{\hspace{3em}}ccc}
    \toprule
    \multirow{2}{*}{\(k\)} & \multirow{2}{*}{\(\hat\nu\)} & \multirow{2}{*}{\(\widehat{\pi}_{k}(\tolerancebandshort)\)} & \multirow{2}{*}{\(\widehat{\pi}_{1}(\tolerancebandshort)\)} & \multirow{2}{*}{ES} & \multicolumn{3}{c}{$\mathrm{ES}$} \\
    \cmidrule(lr){6-8}
    & & & & & $d^{\mathrm{in}}$ & $d^{\mathrm{out}}$ & $\mathrm{sp}$ \\
    \midrule
    \(0.5\)  & \(6.67\) & \(0.975\) & \(0.985\) & \(0.56\) & \(6.2\) & \(5.7\) & \(20.6\) \\
    \(0.75\) & \(4.73\) & \(0.95\)  & \(0.96\)  & \(0.78\) & \(5.9\) & \(5.7\) & \(20.9\) \\
    \(1\)    & \(3.62\) & \(0.92\)  & \(0.92\)  & \(1.03\) & \(6.0\) & \(5.5\) & \(20.4\) \\
    \(1.5\)  & \(2.77\) & \(0.94\)  & \(0.575\) & \(1.31\) & \(5.9\) & \(5.7\) & \(20.1\) \\
    \(3\)    & \(0.19\) & \(0.91\)  & \(0.305\) & \(2.02\) & \(6.3\) & \(5.4\) & \(20.1\) \\
    \(4\)    & \(0\)    & \(0.975\) & \(0.335\) & \(2.09\) & \(6.1\) & \(5.7\) & \(19.3\) \\
    \bottomrule
  \end{tabular}
  }
\end{table}

As the band widens (\cref{tab:eps-sens}), the estimated multiplier decreases monotonically, reaching
\(\hat\nu=0\) at \(4\hat\varepsilon\); conversely, narrower bands require a stronger barrier. In all cases, the fitted-band coverage remains close to or above the target \(1-\delta\).

The main effect of \(\varepsilon\) is therefore on concentration around the observed network. As \(k\) increases, coverage of the calibrated band \(\widehat{\pi}_{1}(\tolerancebandshort)\) decreases and the energy score on the sufficient statistics increases. By contrast, the coefficient estimates remain within approximately one quarter of a standard error of the main fit, and the energy scores for in-degree, out-degree, and shared-partner distributions remain nearly unchanged. Thus, \(\varepsilon\) controls how tightly \algo concentrates probability mass around \(\obsgraph\), while having little effect on parameter interpretation or on distributions not directly constrained by the model.

This experiment isolates sensitivity to the radius \(\varepsilon\): the scale vector \(\scalevec\), reference distribution, target level \(\delta\), and smoothing parameter \(\rho\) are held fixed. Sensitivity to these additional design choices is not assessed here.

\section{Robustness to the Random Seed}\label[appendix]{app:robustness}

\begin{table}[!h]
  \centering
  \caption{Robustness of the \algo coefficient estimates across five
  independent seeds. Values report the mean and standard deviation (in parentheses).}
  \label{tab:robust-coef}
  \resizebox{\columnwidth}{!}{
      \begin{tabular}{lrlrlr}
      \toprule
      \multicolumn{2}{c}{\textsc{Spec}~1} & \multicolumn{2}{c}{\textsc{Spec}~2} & \multicolumn{2}{c}{\textsc{Spec}~3} \\
      \cmidrule(lr){1-2}\cmidrule(lr){3-4}\cmidrule(lr){5-6}
      $|E|$ & $-6.004 (0.0016)$
        & $|E|$ & $-3.359 (0.0039)$
        & $|E|$ & $-2.321 (0.0038)$ \\
      $H_{\mathrm{GroupC}}$ & $1.553 (0.0032)$
        & $\operatorname{rec}$ & $0.548 (0.0021)$
        & $\operatorname{rec}$ & $0.812 (0.0023)$ \\
      $F^{\mathrm{GroupC}}_1$ & $0.002 (0.0012)$
        & $T_\gamma$ & $1.189 (0.0004)$
        & $T_\gamma$ & $0.740 (0.0018)$ \\
      $F^{\mathrm{GroupC}}_2$ & $0.109 (0.0012)$
        & $C_{\mathrm{use}}$ & $0.828 (0.0014)$
        & $D^{\mathrm{out}}_\phi$ & $-2.369 (0.0044)$ \\
      $T_\gamma$ & $2.242 (0.0016)$
        & $C_{\mathrm{team}}$ & $0.894 (0.0008)$
        & $D^{\mathrm{in}}_\phi$ & $0.298 (0.0066)$ \\
      & & $C_{\mathrm{floor}}$ & $0.026 (0.0025)$
        & $C_{\mathrm{use}}$ & $0.808 (0.0009)$ \\
      & & $C_{\mathrm{metric}}$ & $-0.0428 (0.0000)$
        & $C_{\mathrm{team}}$
        & $1.148 \pm 0.0017$ \\
      & & &
        & $C_{\mathrm{floor}}$ & $-0.063 (0.0023)$ \\
      & & &
        & $C_{\mathrm{metric}}$ & $-0.0527 (0.0000)$ \\
      \bottomrule
      \end{tabular}
    }
\end{table}

\begin{table}[!h]
  \centering
  \small
  \caption{Robustness of \algo across five seeds.
  Values report mean (std). $\widehat{\pi}^{(W)}(\mathcal B)$ is the
  mean band concentration over the final $W$ iterations, and
  $\widehat{\pi}_{\hat\theta,\hat\nu}(\mathcal B)$ is the concentration
  estimated from simulations at the final fitted parameters.}
  \label{tab:robust-fit}
  \setlength{\tabcolsep}{4pt}
  \begin{tabular*}{\columnwidth}{@{\extracolsep{\fill}}lccccc@{}}
  \toprule
    & $\hat\nu$
    & $\widehat{\pi}^{(W)}(\mathcal B)$
    & $\widehat{\pi}_{\hat\theta,\hat\nu}(\mathcal B)$
    & $R$
    & ES \\
  \midrule
  \textsc{Spec}~1 & 2.69 (0.45) & 0.90 (0.01) & 0.83 (0.10) & 1.21 (0.71) & 0.85 (0.25) \\
  \textsc{Spec}~2 & 3.85 (0.19) & 0.90 (0.00) & 0.94 (0.01) & 0.31 (0.14) & 0.84 (0.02) \\
  \textsc{Spec}~3 & 3.62 (0.07) & 0.89 (0.01) & 0.95 (0.02) & 0.29 (0.06) & 1.00 (0.02) \\
  \bottomrule
  \end{tabular*}
\end{table}

To assess the sensitivity of the results to algorithmic randomness, we repeated
the estimation of each specification using five different random seeds, while
keeping the band and all other estimation settings fixed
(\Cref{tab:config-ext}), except for the initialization of \textsc{Spec}~3
described below. Each run starts from the MPLE and, for \textsc{Spec}~1,
repeats both estimation stages.\footnote{The main run of
\textsc{Spec}~3 was initialized from a preliminary estimate, whereas the five
robustness runs start from the MPLE.}
The calibrated band, i.e., $\varepsilon$ and $\scalevec$, is kept fixed across
runs. After estimation, we independently sample from the exact model at the
fitted parameters using the same settings as in the main evaluation.

We distinguish two estimates of band concentration: $\widehat{\pi}^{(W)}(\tolerancebandshort)$ 
is the mean estimated concentration over the final $W$ estimation iterations, and
$\widehat{\pi}_{\hat{\thetavec},\hat\nu}(\tolerancebandshort)$ is the
concentration obtained from the independent post-estimation simulation of the
exact model.

\Cref{tab:robust-coef} reports the coefficient estimates across seeds.
The estimates are highly stable: the standard deviation of every coefficient
is below $0.007$, and the mean estimates differ from those of the main runs by
at most $0.02$. For \textsc{Spec}~3, the runs initialized from the MPLE also
converge to essentially the same estimates as the main run, despite its
different initialization.

\Cref{tab:robust-fit} reports $\hat\nu$,
$\widehat{\pi}^{(W)}(\tolerancebandshort)$,
$\widehat{\pi}_{\hat{\thetavec},\hat\nu}(\tolerancebandshort)$, $R$, and $\mathrm{ES}$.
For \textsc{Spec}~2 and~3, all quantities vary little across seeds, and the
post-estimation concentrations are $0.94$ (std $0.01$) and $0.95$ (std $0.02$),
respectively, exceeding the target $1-\delta$ in every run.

For \textsc{Spec}~1, both $\hat\nu$ and the post-estimation concentration vary
more: $\hat\nu$ has mean $2.69$ (std $0.45$), while
$\widehat{\pi}_{\hat{\thetavec},\hat\nu}(\tolerancebandshort)$ has mean $0.83$
(std $0.10$), ranging from $0.69$ to $0.95$. This is lower on average than the
concentration of approximately $0.90$ measured during estimation. The
difference is consistent with the bimodality discussed in
\Cref{sec:ukFaculty}: the chains used during estimation restart from the
observed graph at every iteration and rarely transition to the sparse mode,
whereas the longer post-estimation chains are more likely to reveal its
probability mass. Since transitions between the two modes are rare and the
mass of the sparse mode is sensitive to $\nu$, its estimation is also more
variable across seeds. The main-run concentration of $0.81$ is close to the
five-run average and, in every run, remains substantially above the $7\%$
concentration of the standard ERGM.

\section{Additional Results}\label[appendix]{sec:add-results}

\subsection{Distributions of the Sufficient Statistics}\label{sec:suff-stats}

\begin{figure}[htb]
  \centering
  \includegraphics[width=\linewidth]{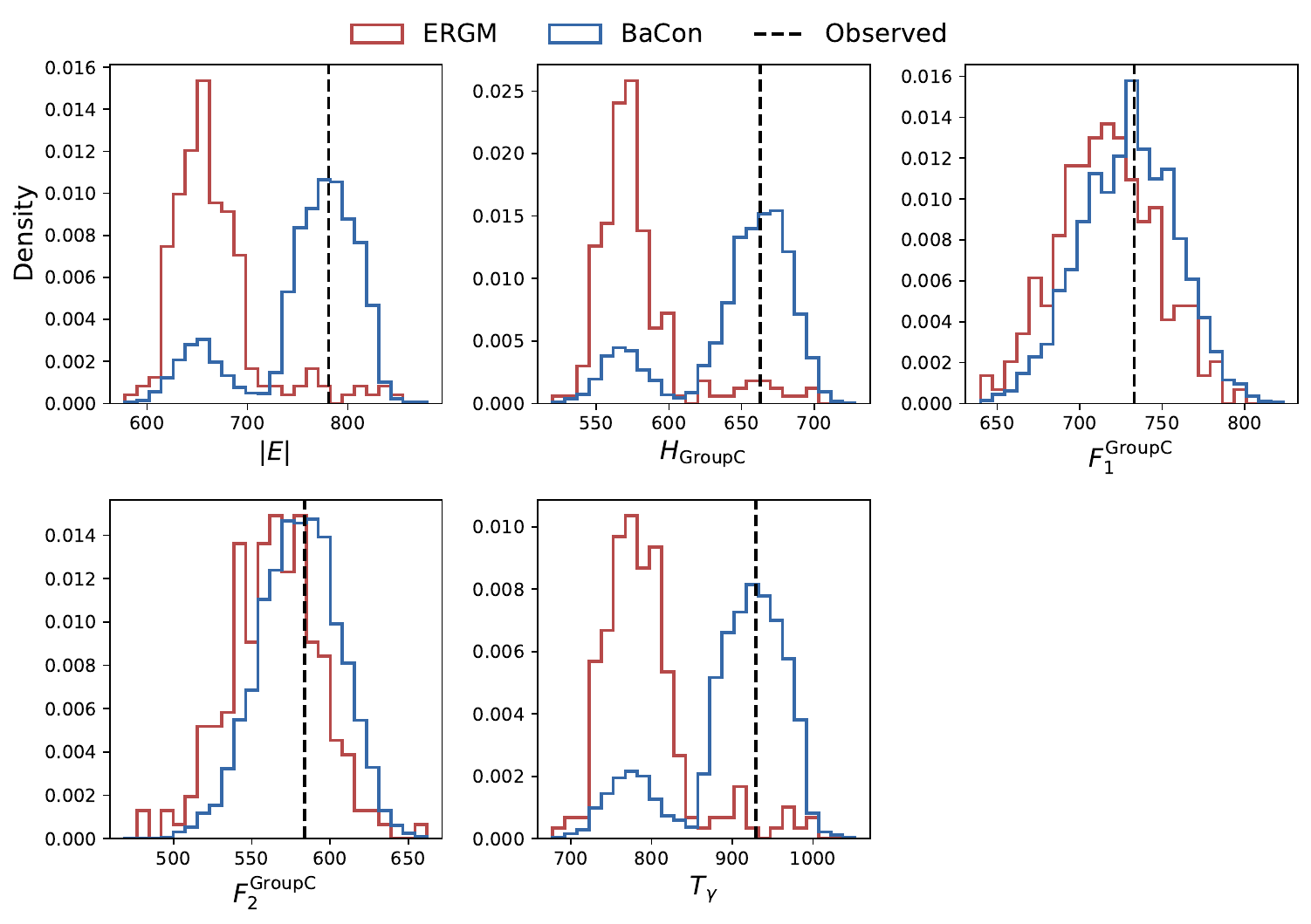}
  \caption{\textsc{Spec}~1. Distribution of the sufficient statistics under
  \algo and ERGM, and corresponding observed statistics.}%
  \label{fig:stats_all_spec_1}
\end{figure}

\begin{figure}[!ht]
  \centering
  \includegraphics[width=\linewidth]{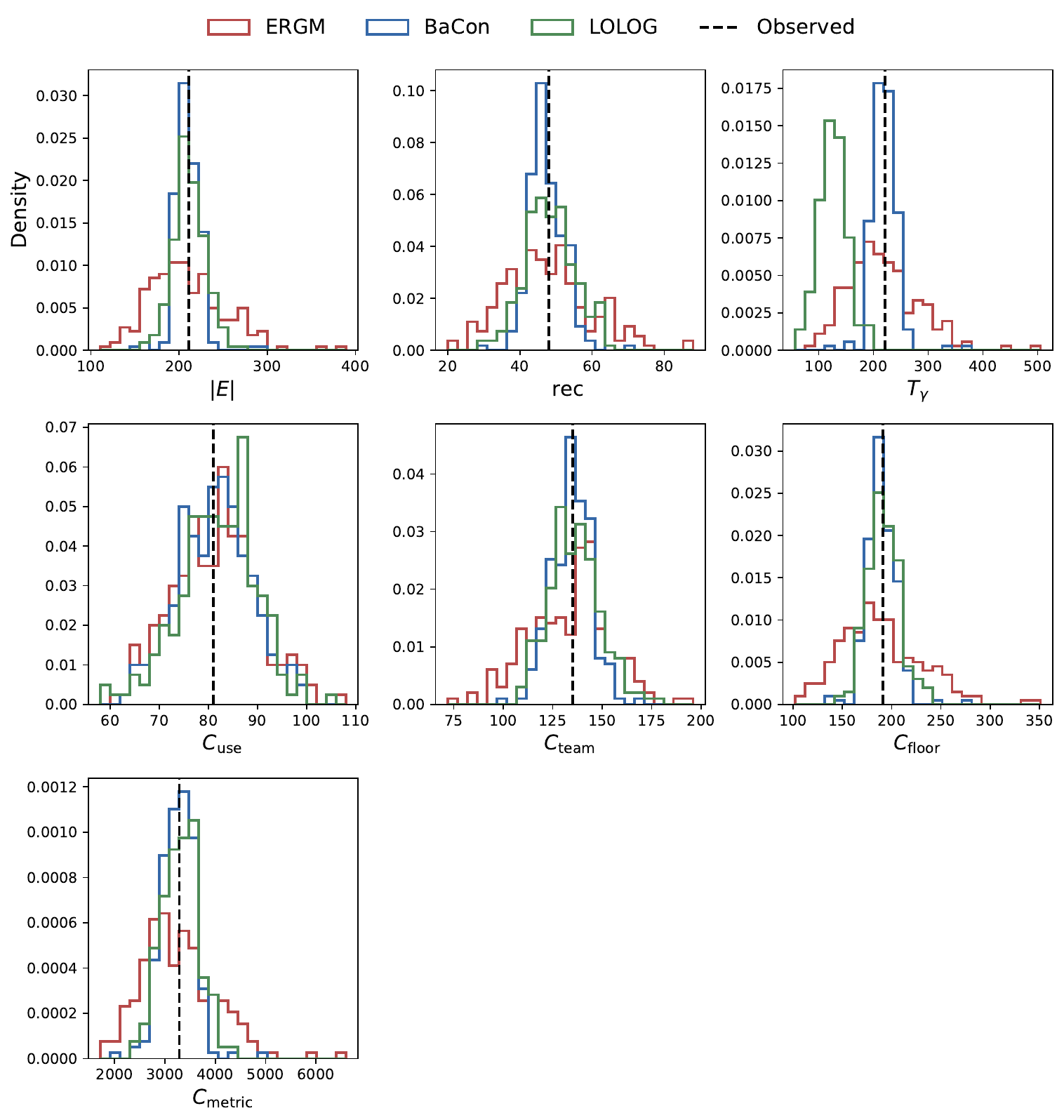}
  \caption{\textsc{Spec}~2. Distribution of the sufficient statistics under
  \algo, ERGM, and LOLOG, and corresponding observed statistics.}%
  \label{fig:stats_all_spec_2}
\end{figure}

\begin{figure}[!ht]
  \centering
  \includegraphics[width=\linewidth]{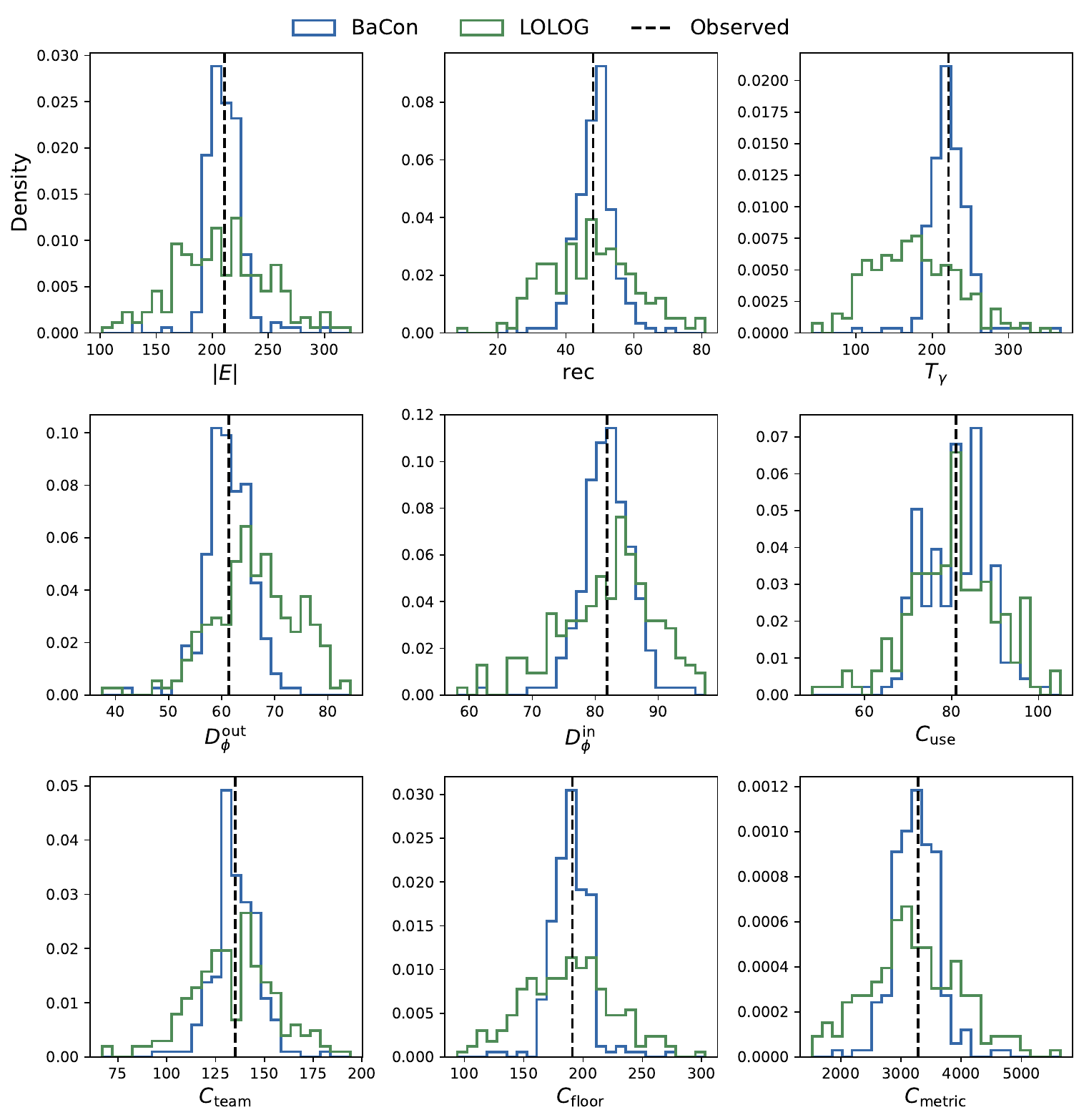}
  \caption{\textsc{Spec}~3. Distribution of the sufficient statistics under
  \algo and LOLOG, and corresponding observed statistics.}%
  \label{fig:stats_all_spec_3}
\end{figure}

Figures~\ref{fig:stats_all_spec_1}--\ref{fig:stats_all_spec_3} provide the full
goodness-of-fit distributions of the model statistics.

For \textsc{Spec}~1, the difference between ERGM and \algo is
already visible at the level of the sufficient statistics. The
\algo simulations are concentrated around the observed values,
whereas the ERGM distributions are substantially displaced for, in
particular, $|E|$, $T_\gamma$, and $H_{\mathrm{GroupC}}$. Thus, although
standard ERGM estimation reports convergence, the resulting distribution
does not make the observed network typical. In fact, most ERGM simulations
fall in a sparse mode, while \algo moves most of the probability
mass back toward the observed configuration. The correction is not
complete: a smaller sparse mode remains under \algo, explaining
why its coverage remains below the nominal level.

For \textsc{Spec}~2, both ERGM and \algo are correctly centered:
their simulated distributions are aligned with the observed values for the
model statistics. Their dispersion, however, is markedly different.
The ERGM simulations are substantially more spread out, whereas
\algo concentrates probability mass around the observed
configuration. This illustrates why right moment matching alone does not
guarantee typicality: ERGM is centered but places only a small fraction of
its simulations close to the observation, while the \algopref constraint
substantially increases this concentration.

LOLOG is also centered on most statistics, but appears shifted on
$T_\gamma$. This discrepancy is due to the different definition of the
directed shared-partner statistic: LOLOG uses cyclic shared partners,
whereas the ERGM and \algo specifications use the transitive
version. It should therefore not be interpreted as a failure of LOLOG to
reproduce one of its own sufficient statistics. When evaluated using its
own cyclic statistic, LOLOG is well centered, although its simulations
remain less concentrated than those of \algo.%

For \textsc{Spec}~3, \algo again produces distributions tightly
concentrated around the observed statistics. LOLOG is generally much more
dispersed, with the two models behaving similarly only for
$C_{\mathrm{use}}$. This difference is consistent with the in-model
diagnostics: \algo attains high coverage, whereas the LOLOG
simulations remain widely dispersed even when evaluated on the statistics
of their own specification.

The comparison must nevertheless be interpreted with some care because
the two models do not use identical sufficient statistics. In particular,
LOLOG replaces the geometrically weighted in- and out-degree terms with
second- and third-order in- and out-stars and again uses cyclic rather than
transitive shared partners. 

\subsection{Adequacy}\label{sec:adequacy}

\begin{figure}[!ht]
  \centering
  \begin{subfigure}{\linewidth}
    \centering
    \includegraphics[width=.95\linewidth]{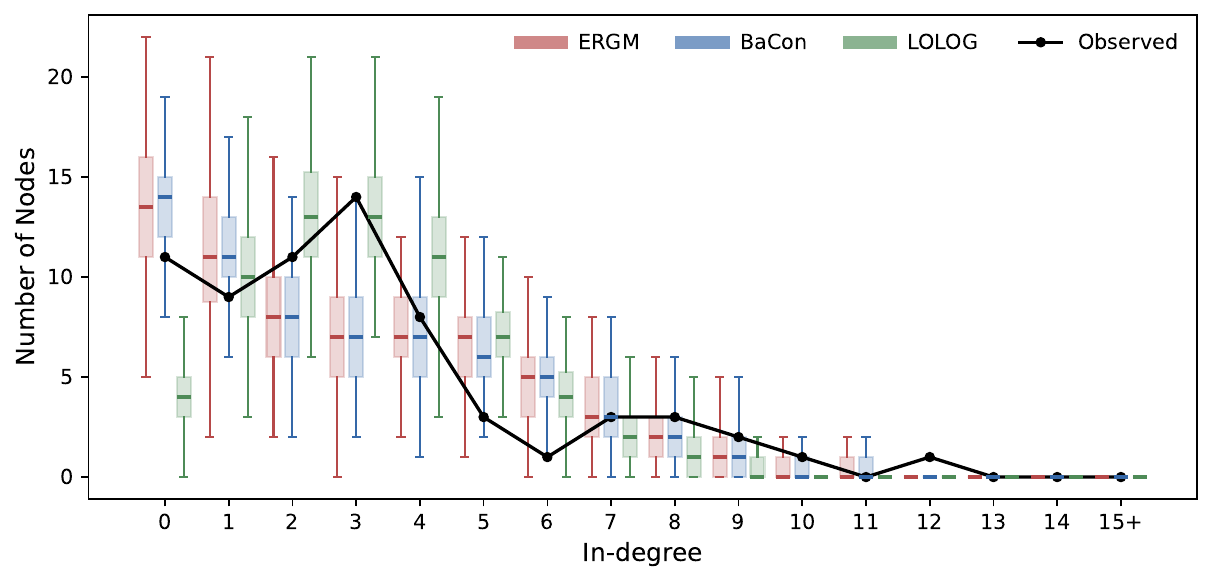}
    \includegraphics[width=.95\linewidth]{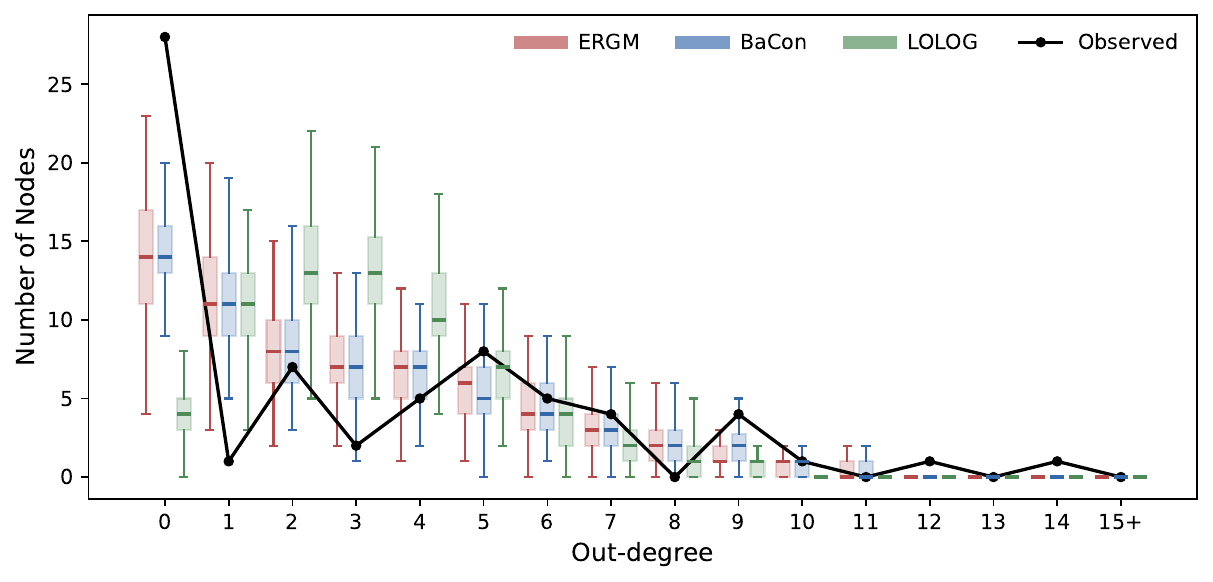}
    \caption{\textsc{Spec}~2}%
    \label{fig:inoutdeg_spec_2}
  \end{subfigure}
  \begin{subfigure}{\linewidth}
    \centering
    \includegraphics[width=.95\linewidth]{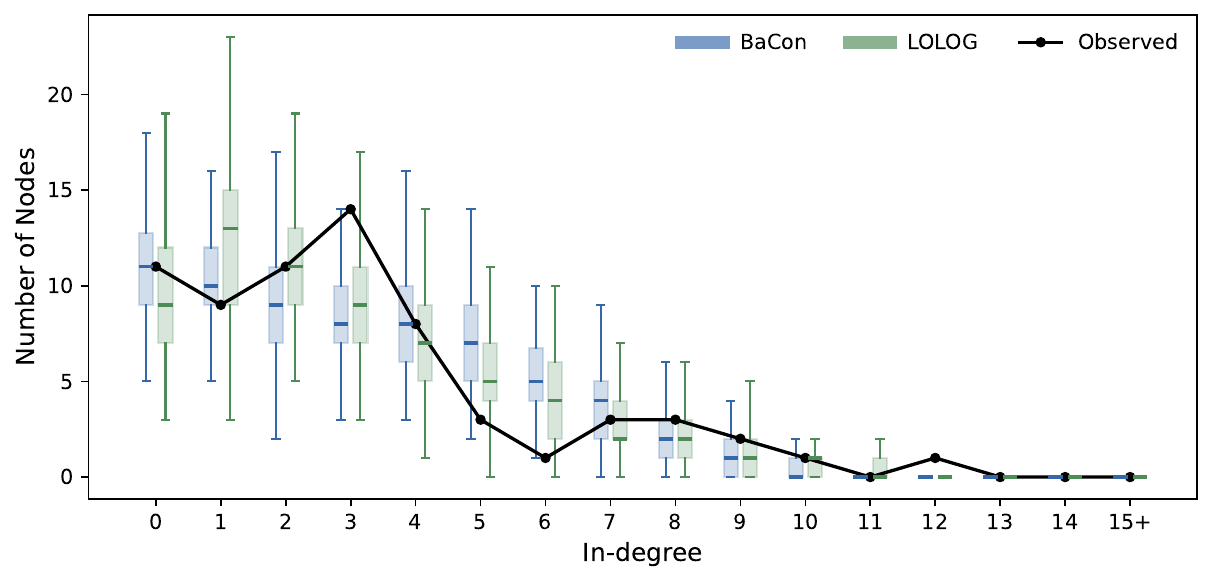}
    \includegraphics[width=.95\linewidth]{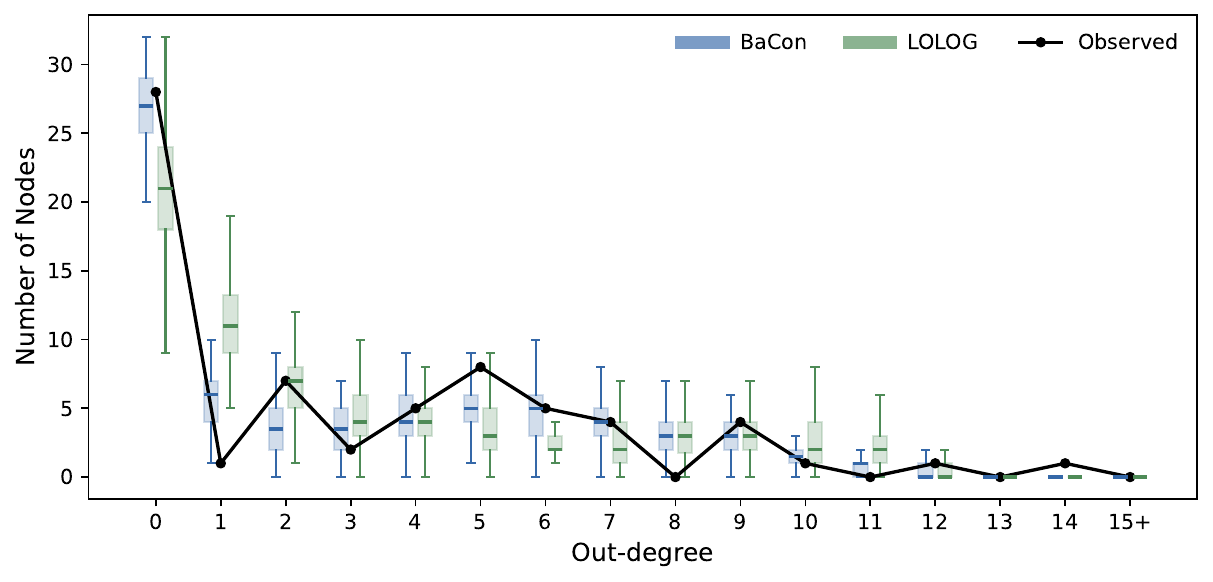}
    \caption{\textsc{Spec}~3}%
    \label{fig:inoutdeg_spec_3}
  \end{subfigure}
  \caption{Distribution of in-degree (top) and out-degree (bottom) under the
  various models, together with the observed values. Degrees are shown
separately up to 14, while all degrees greater than or equal to 15 are grouped
in the final bin.}%
  \label{fig:inoutdeg_spec_2_3}
\end{figure}

\enlargethispage{\baselineskip}

\Cref{fig:inoutdeg_spec_2} provides a more detailed view of the
out-of-model adequacy results reported in the main paper. For the in-degree
distribution, the three models behave similarly in the upper part of the
distribution: they tend to overestimate the number of nodes with degree
5--6 and underestimate the number with degree at least 7. The main
differences occur at smaller degrees. LOLOG is closer to the observation
around degrees 1--3, whereas ERGM and \algo are closer at
degrees 0 and 4. These differences largely compensate when aggregated,
which is consistent with the very similar in-degree energy scores and
$\ell_1$ distances reported in the paper.

For the out-degree distribution, ERGM and \algo are again
closely aligned and are generally closer to the observed distribution than
LOLOG for degrees 0--5. Above degree 6, the three models behave similarly,
without a systematic direction of error: depending on the degree, they
alternately over- or underestimate the observed counts. The largest common
shortcoming is at degree zero, where the observed network contains many
more nodes with no outgoing ties than the simulated networks. This explains
why the out-degree discrepancy remains considerably larger than for the
in-degree and motivates the degree terms introduced in
\textsc{Spec}~3.

Overall, the close agreement between ERGM and \algo on these
out-of-model distributions is consistent with their closely aligned
coefficient estimates. The \algopref constraint substantially improves the fit of
the sufficient statistics and the typicality of the observed network, but it
does not directly constrain the full degree distributions. Consequently,
there is no reason to expect the improvement in in-model concentration to
translate automatically into an improvement in these out-of-model
quantities. This is also consistent with the maximum-entropy construction:
subject to the imposed moment and coverage constraints, \algo
otherwise preserves as much uncertainty as possible rather than targeting
additional structural features not included in the specification.

\enlargethispage{\baselineskip}

For \textsc{Spec}~3, LOLOG and \algo reproduce the degree distributions similarly overall, with their simulated means generally close to the observed counts. The agreement is particularly good at the smaller degrees, where the observed distribution is most concentrated. LOLOG shows greater variability in these bins, whereas \algo produces more concentrated distributions around the observed values. At larger degrees, the two models behave comparably, and neither exhibits a systematic bias.

The improvement over \textsc{Spec}~2 is most evident for the out-degree distribution: the added degree terms allow \algo to reproduce much more closely the large number of nodes with very small out-degree, which was the main deficiency of the previous specification. Overall, the figure confirms that the two approaches achieve similar average adequacy, while \algo tends to generate less variable degree profiles, especially in the low-degree region.

\end{document}